\documentclass[%
superscriptaddress,
showpacs,preprintnumbers,
nofootinbib,
 amsmath,amssymb,
 aps, 
twocolumn,
prl,
]{revtex4}
\usepackage{graphicx}
\usepackage{amssymb}
\usepackage{amsbsy}
\usepackage{amsmath}
\usepackage{chngcntr}
\usepackage[export]{adjustbox}
\usepackage{color}
\usepackage{bbm, dsfont}

\newcommand {\fabsq}[1] {\left| #1 \right|^2}
\newcommand {\fabs}[1] {\left| #1 \right|}
\def\ra{\rangle}
\def\la{\langle}

\newcommand{\beqa}{\begin{eqnarray}}
\newcommand{\eeqa}{\end{eqnarray}}
\newcommand{\beq}{\begin{equation}}
\newcommand{\eeq}{\end{equation}}

\newcommand{\ket}[1]{|#1\rangle}

\newcommand{\braket}[2]{\langle#1|#2\rangle}

\newcommand{\suppA}{(see Appendix A)}
\newcommand{\suppB}{(see Appendix B)}
\newcommand{\insuppB}{in Appendix B}
\newcommand{\seesuppC}{see Appendix C}

\begin{document}
\title{Quantum-optical implementation of non-Hermitian potentials for asymmetric scattering}
\author{A. Ruschhaupt}
\affiliation{Department of Physics, University College Cork, Ireland}
\author{A. Kiely}
\affiliation{School of Physics, University College Dublin, Belfield 4, Ireland}
\author{M. A. Sim\'on}
\affiliation{Departamento de Qu\'{\i}mica F\'{\i}sica, UPV/EHU, Apdo
644, 48080 Bilbao, Spain}
\author{J. G.  Muga}
\affiliation{Departamento de Qu\'{\i}mica F\'{\i}sica, UPV/EHU, Apdo
644, 48080 Bilbao, Spain}
\begin{abstract}
Non-Hermitian, one-dimensional  potentials which are also non-local, 
allow for scattering asymmetries, namely, asymmetric transmission or reflection responses to the incidence of a particle from left or right.
The  symmetries of the potential 
imply selection rules for transmission and reflection. In particular, parity-time (PT)  
symmetry or the symmetry  of any local potential do not allow for asymmetric transmission.  
We put forward a feasible  quantum-optical implementation  
of non-Hermitian, non-local, non-PT potentials to implement different scattering asymmetries, including transmission
asymmetries.    
\end{abstract}
\maketitle
{\it Introduction.}
The asymmetric response of diodes, valves, or rectifiers to input direction is of paramount importance in many different fields and technologies, from hydrodynamics  to microelectronics,  as well as in biological systems. We  expect a wealth of applications of such response asymmetries also in the microscopic quantum realm, in particular in  circuits or operations carrying or processing quantum information with moving atoms.  
So far  devices such as Maxwell demons, which  let atoms pass one way, have  been instrumental, first as ideal devices to  understand the second law \cite{Maxwell1875,Maxwell1990}, and also  as practical sorting devices \cite{Ruschhaupt2004_diode,Raizen2005,Dudarev2005,Ruschhaupt2006a,Ruschhaupt2006b,Ruschhaupt2006c,Ruschhaupt2007,Raizen2009,Jerkins2010}.    
      
Asymmetric transmission and reflection 
probabilities for one-dimensional (1D) particle scattering off  a potential center are not possible if the Hamiltonian is Hermitian \cite{Muga2004,Mosta2018}. 
Non-Hermitian (NH) Hamiltonians representing effective interactions have a long history in nuclear, atomic, and molecular physics, and have become common in optics, where wave equations in waveguides could simulate  Schr\"odinger equations \cite{Ruschhaupt2005,optics1,optics2}. 
Non-Hermitian Hamiltonians constructed by analytically continuing Hermitian ones may be useful tools to find resonances \cite{Moiseyev}. 
They can also be set phenomenologically, e.g. to describe gain and loss 
\cite{Ruschhaupt2005},  
or be found as effective Hamiltonians for a subspace from a Hermitian Hamiltonian of a larger system 
by projection \cite{Feshbach1958,Ruschhaupt2004,Muga2004}.  

Much of the recent interest in Non-Hermitian Hamiltonians focuses on  parity-time (PT) symmetric Hamiltonians \cite{Bender1998,znojil}  because of their spectral properties and useful applications, mostly in optics  \cite{optics1,optics2,Longhi2014}, but  alternative symmetries are also being studied \cite{Nixon2016a,Nixon2016b,Chen2017,Ruschhaupt2017,Simon2018,Simon2019,Alana2020,Bernard2002,Kawabata2019}. Symmetry operations 
on NH Hamiltonians can be systematized into group structures \cite{Ruschhaupt2017,Simon2019,Alana2020}.  In particular for 
1D particle scattering off a potential center, the different Hamiltonian symmetries imply 
selection rules for asymmetric transmission and reflection \cite{Ruschhaupt2017,Simon2019}.
Whereas hermiticity does not allow for any asymmetry in transmission and reflection probabilities,   PT symmetry or 
the symmetry of local potentials, technically ``pseudohermiticy with respect to time reversal'' \cite{Ruschhaupt2017},    
do not allow for  asymmetric transmission \cite{Muga2004,Mosta2018}, see symmetries II, VI,  and VII in table \ref{condi}. 
(Here a ``local potential'' is defined as one whose only non-zero elements in coordinate representation are diagonal, whereas a non-local one has  non-zero nondiagonal elements.) 
Thus  non-local, non-PT, and non-Hermitian potentials are needed to implement a rich set of scattering
asymmetries, and in particular asymmetric transmission.  

In this paper we put forward a physical realization of  effective NH, non-local  Hamiltonians which do not posses PT symmetry.  
Non-local potentials for asymmetric scattering had been constructed as mathematical models \cite{Ruschhaupt2017}, but a physical implementation  had been so far elusive. 
Using Feshbach's projection technique it is found that the 
effective potentials for a ground-state atom crossing a laser beam in a region of space are generically non-local and non-Hermitian. Shaping the spatial-dependence of the, generally complex, Rabi frequency, and selecting a specific laser detuning allows us to produce different potential symmetries and asymmetric scattering effects, including asymmetric transmission.     

After a lightning review of Hamiltonian symmetries and the corresponding scattering selection rules,   
we shall  explain how to generate different NH symmetries in a quantum optical setting of an atom impinging on a laser illuminated region. Finally we provide specific examples with different asymmetric scattering responses.  

{\it Symmetries of Scattering Hamiltonians.\label{ssh}}
We consider one-dimensional  scattering Hamiltonians  $H=H_0+V$, where $H_{0}=p^{2}/(2m)$ is the kinetic energy for a particle of mass $m$,  $p$ being the momentum, and
$V$ is the  potential, which is assumed to decay fast enough on both sides so that $H$ has a continuous spectrum and scattering eigenfunctions. These eigenfunctions may be chosen so that asymptotically, i.e., far from the potential center,  they are superpositions of an incident plane wave and a reflected plane wave on one side, and a transmitted plane wave on the other side. Reflected and transmitted waves include corresponding amplitudes, whose squared-modulii 
(scattering coefficients hereafter) sum to one for Hermitian potentials. Instead, NH potentials 
may produce  absorption or gain.

There are eight different symmetries that $H$ could fulfill, see table \ref{condi}, 
with the forms 
\begin{eqnarray}
AH&=&HA, \label{commutation}
\\
AH&=&H^{\dagger}A,
\label{pseudohermiticity}
\end{eqnarray}
where $A$ is a unitary or antiunitary operator in the Klein four-group $K_{4}=\lbrace 1,\Pi,\theta,\Pi\theta \rbrace$ \cite{Ruschhaupt2017}.
Relation (\ref{pseudohermiticity}) is called here $A$-pseudohermiticity of $H$ \cite{Mosta2010,Ruschhaupt2017}.  
The operators $\Pi$, $\theta$ and $\Pi\theta$ are parity, time reversal, 
and the consecutive (commuting) application of both operators. 
Acting on position eigenvectors $|x\rangle$,  
$\Pi c|x\rangle =c|-x\rangle$, and $\theta c |x\rangle = c^* |x\rangle$, for any  complex number $c$. 

The eight symmetries  may be regarded as the invariance of the Hamiltonian with respect to
eight symmetry operations that form the Abelian group E8 \cite{Simon2019}. They are all operations that can be done by inversion, transposition, complex conjugation, and their combinations. 
Making use of generalized unitarity relations and  the relations implied by the symmetries on $S$-matrix elements, 
the transmission and reflection amplitudes for right and left incidence, $T^r$, $R^r$ and $T^l$, $R^l$, can be  related
to each other,
as well as their modulii \cite{Ruschhaupt2017}.  ``Right and left  incidence'' are here shorthands for ``incidence {\it from} the right'', 
and ``incidence { from} the left'', respectively.   

The possible asymmetric responses are allowed or forbidden, according to selection rules,  by the symmetries of the Hamiltonian. 
If we impose that the transmission and reflection coefficients  have only 0 or 1 values,  a convenient reference scenario for  devices intended to manage quantum-information applications, six possible scattering asymmetries may be identified \cite{Ruschhaupt2017}, see table \ref{table2}. It is useful to label them according to the response  to incidence from the left/right. The possible responses are encoded in the  letters ${\cal{A}}$, 
for ``absorption'', and ${\cal{T}}$ and ${\cal{R}}$ for ``transmission'' and ``reflection'', separated by ``$/$''. The   
letters on the left of $/$ are for left incidence, and the ones on the right are for right incidence. For example ${\cal T/A}$ means transmission for left incidence and absorption for right
incidence. From the selection rules \cite{Ruschhaupt2017}, it is possible to determine which symmetries allow for a given device, see table \ref{table2}.

\begin{table}[t]

\caption{Conditions leading to  specific symmetries in the potential (\ref{effpot}). A given symmetry also implies others, see the last column.\label{condi}}
\hspace*{-0.1cm}
\scalebox{.94}{
\begin{tabular}{lcc}
\hline\hline
			Symmetry& Conditions & Implies 
			\\
			\hline
			(I)\;$1H=H1$ &   none & - 
			\\
			(II)\;$1H=H^\dagger 1$ &  $q=-q^{*}$ (i.e. $\operatorname{Re}q=0$) & I 
			\\
			(III)\;$ \Pi H=H\Pi$ &  $\Omega(x)=e^{i\phi}\Omega(-x)$ & I 
			\\
			(IV) $\Pi H=H^\dagger \Pi$ &  $q=-q^{*}$ \& $\Omega(x)=e^{i\phi}\Omega(-x)$ & III,\! II,\! I 
			\\
			(V) $\Theta H=H\Theta$ &  $q=-q^{*}$ \& $\Omega(x)=e^{i\phi}\Omega(x)^*$ & {\small VI,\! II,\! I} 
			\\
			(VI) $\Theta H=H^\dagger\Theta$ &  $\Omega(x)=e^{i\phi}\Omega(x)^*$ & I 
			\\
			(VII) $\Theta\Pi H=H\Theta \Pi$ &  $q=-q^{*}$ \& $\Omega(x)=e^{i\phi}\Omega(-x)^*$ & VIII,\! II,\! I 
			\\
			(VIII)\,$\Theta\Pi H=H^\dagger \Theta \Pi$ &  $\Omega(x)=e^{i\phi}\Omega(-x)^*$  & I
			\\
\hline\hline			 
\end{tabular}
}
\end{table}

\begin{table}[t]
\caption{Device types for  transmission and/or reflection asymmetry in the first row.
The second row gives the corresponding symmetries  that allow
each device.
\label{devices}
\vspace*{.0cm}}
\label{table2}
\centering
\begin{ruledtabular}
\scalebox{0.90}{
\begin{tabular}{cccccc}
$\cal{TR/A}$ & $\cal{T/R}$ & $\cal{T/A}$ & $\cal{TR/R}$ & $\cal{R/A}$ & $\cal{TR/T}$ \\
I            & I           & I,VIII      & I,VIII       & I,VI        & I, IV, VI, VII 
\end{tabular}}
\end{ruledtabular}
\end{table}

{\it Effective non-local potential for the ground state of a two-level atom.\label{enl}}
The key task is now to physically realize some of the potential and device types described in the previous section. We start with a two-level atom with ground level $|1\ra$ and excited state $|2\ra$ impinging onto a laser illuminated region. For a full account of the model and further references see 
\cite{Ruschhaupt2009}. The motion is assumed one dimensional, either because the atom is confined in a waveguide or because the direction $x$ is uncoupled to 
the others.
We only account explicitly for atoms before the first spontaneous emission in the wavefunction  
\cite{Hegerfeldt1996,Damborenea2002,Navarro2003}.  
If the excited atom emits a spontaneous photon it disappears from the coherent wavefunction ensemble. 
We assume that no resetting into the ground state occurs. The physical mechanism 
may be an irreversible decay into a third level  \cite{Oberthaler1996}, or atom ejection from the waveguide or the privileged 1D direction due to  random recoil  \cite{Streed2006}.   
The state ${\bf\Phi}_k=\left(\begin{smallmatrix}\phi_k^{(1)}\\\phi_k^{(2)}\end{smallmatrix}\right)$
for the atom before the first spontaneous emission impinging with wavenumber $k$  
in a laser adapted  interaction picture, 
obeys, after applying the rotating wave approximation,  an effective stationary Schr\"odinger equation
with a time-independent Hamiltonian \cite{Ruschhaupt2004,Ruschhaupt2009}
${\cal H}{\bf\Phi}_k(x)=E{\bf\Phi}_k(x)$,
where
\beqa
{\cal H}&=&K{\bf 1}+ {\cal V}=\frac{1}{2m}\left(
{{p}^2\atop 0}{0\atop {p}^2}\right)+ {\cal V}(x),
\\
{\cal V}(x) &=&
\frac{\hbar}{2}\left(
{0\atop \Omega(x)^*}\;\;\;\;
{\Omega(x)\atop -(2\Delta+i\gamma)}
\right).
\eeqa
We assume perpendicular incidence of the atom on the laser sheet for simplicity, oblique incidence is treated e.g. in  \cite{Ruschhaupt2007,Ruschhaupt2009}.  
Here $E=\hbar^2 k^2/2m$ is the energy, and 
$\Omega(x)$ is the position-dependent, on-resonance Rabi frequency, where real and imaginary parts may be controlled independently 
using two  laser field quadratures  \cite{Zhang2013};  
$\gamma$ is the inverse of the life time of the excited state;
$\Delta=\omega_{L}-\omega_{12}$
is the detuning (laser angular frequency minus the atomic transition 
angular frequency $\omega_{12}$); 
$K={p}^2/2m$ is the kinetic energy,
${p}=-i\hbar\partial/\partial x$; 
and ${\bf 1}=|1\ra\la 1|+|2\ra\la 2|$ is the unit operator 
for the internal-state space.
Complementary projectors 
$P=|1\ra\la 1|$ and $Q=|2\ra\la 2|$
are defined to select ground and excited state components.  
Using the partitioning 
technique \cite{Feshbach1958,Feshbach1962,Levine1969},
we find for the ground 
state amplitude $\phi_k^{(1)}$ the equation 
\beq\label{effecti}
E\phi_k^{(1)}(x) = K\phi_k^{(1)}(x)+\!
\int\! dy\, \la x,1|{\cal W}(E)|y,1\ra \phi_k^{(1)}(y),
\eeq
where  
$
{\cal W}(E)=P{\cal V}P+P{\cal V}Q(E+i0-Q{\cal H}Q)^{-1}Q{\cal V}P,
$
is generically non local and energy dependent. Specifically, we have now achieved
a physical realization of an effective (in general) non-local, non-Hermitian potential of the form
\beqa
\hspace*{-.3cm}V (x,y) = \la x,1|{\cal W}(E)|y,1\ra = \frac{m}{4} \frac{e^{i|x-y|q}}{i q}
\Omega(x)\Omega(y)^*,
\label{effpot}
\eeqa
where  
$
q=\frac{\sqrt{2mE}}{\hbar}(1+\mu)^{1/2},\;\; 
{\rm Im}\,q\ge 0,
\label{qeq}
$ and 
$
\mu=\frac{2\Delta+i\gamma}{2E/\hbar}.
$
Eq. (\ref{effpot}) is worked out  in momentum representation to do the integral  
using the residue theorem. 
This is a generalized, non-local version of the effective potentials known for the ground state 
\cite{Chudesnikov1991,Oberthaler1996}, which are found from Eq. (\ref{effpot})  in the large $\mu$ limit \cite{Ruschhaupt2004}.   
The reflection and transmission amplitudes $R^{r,l}$ and  $T^{r,l}$ may be calculated directly
using  the potential  (\ref{effpot}) or as  corresponding amplitudes for 
transitions from ground state to ground state in the full two-level theory \suppA.

{\it Possible symmetries of the non-local potential.}
The necessary conditions for the different symmetries of the potential (\ref{effpot}) are outlined in the second column of  table \ref{condi}. Since $\Omega(x)$ does not depend on $q$, symmetries IV, V and VII imply that symmetry II is obeyed as well (Hermiticity). 
Moreover symmetry III (parity) should be discarded for our purpose since it does not allow for asymmetric transmission or reflection
\cite{Ruschhaupt2017}. 
This leaves us with three interesting symmetries to explore: 
VI, which allows for  asymmetric reflection; VIII which allows for asymmetric transmission, and  I, 
which in principle allows for arbitrary asymmetric responses, except for physical limitations imposed by
the two-level model \suppA.

As seen from  table \ref{condi}, $\operatorname{Re}(q)=0$ makes the potential Hermitian so we shall avoid this condition. 
If $\gamma=0$,   $\mu \in \mathbb{R}$. Hence $\mu+1<0$ gives $\operatorname{Re}(q)=0$ and $\mu+1>0$ gives 
$\operatorname{Im}(q)=0$. $\mu+1>0$ amounts to a condition on the detuning compared to the incident energy, namely $\Delta>-E/\hbar$. 
In the following examples we implement potentials with symmetries VIII, VI, and I, with detunings and energies satisfying the condition $\mu+1>0$.   

{\it Design of asymmetric devices.\label{exa}}
We will now apply this method to physically realize non-local potentials of the form \eqref{effpot}. 
We  shall work out  explicitly a ${\cal T/A}$ device with symmetry  VIII, a ${\cal R/A}$ device with symmetry  VI, and a ``half''-${\cal TR/A}$ device with symmetry I. The  ${\cal T/A}$ and the ``half''-${\cal TR/A}$ device have transmission asymmetry so they cannot be built with local or $PT$-symmetric  potentials.
Let us  motivate the effort with some possible applications, relations  and analogies of these devices.   
${\cal T/A}$ and ${\cal R/A}$ are, respectively, transmission and reflection filters. They are analogous to
half-wave electrical rectifiers that either let the signal from one side ``pass'' (transmitted) or change its sign (reflected)
while suppressing the other half signal.  They may play the role of half-rectifiers in atomtronic circuits. 
A ${\cal T/A}$ device allows us, for example, to empty a region of selected particles, letting them go away while not letting particles in. 
The ``atom diode'' devices worked out e.g. in \cite{Ruschhaupt2004_diode,Ruschhaupt2006a,Ruschhaupt2006b,Ruschhaupt2007}
where of type ${\cal R/A}$. As the mechanism behind them was adiabatic, a broad range of momenta with the desired asymmetry
could be achieved. In comparison the current approach is not necessarily adiabatic so it can be adapted to faster processes.   

As for the ``half''-${\cal RT/A}$ device,  it reflects and transmits from one side while absorbing from the 
other side. In an optical analogy an observer from the left perceives it as a darkish mirror. 
An observer from the right ``sees'' the other side because of the allowed transmission 
but cannot be seen from the left since  nothing is transmitted from right to left. Our device is necessarily ``half'' one as 
there cannot be  net probability gain because of the underlying two-level system, and a ``full'' version with both reflection and transmission coefficients equal to one
would need net gain.

\begin{figure}
\begin{center}
\includegraphics[width=0.48\linewidth]{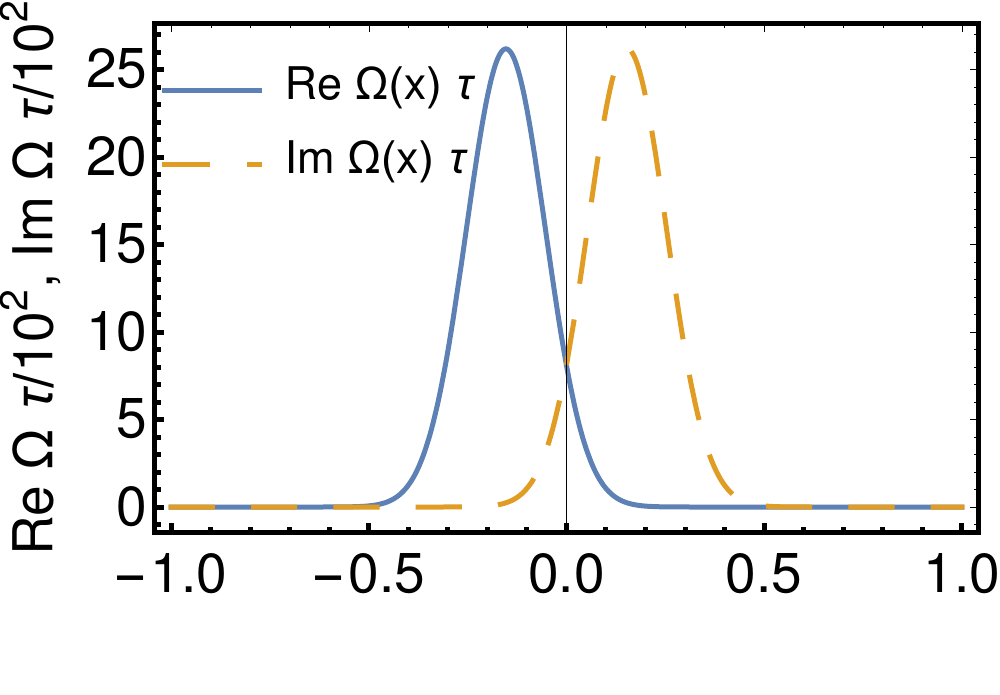}
\includegraphics[width=0.49\linewidth]{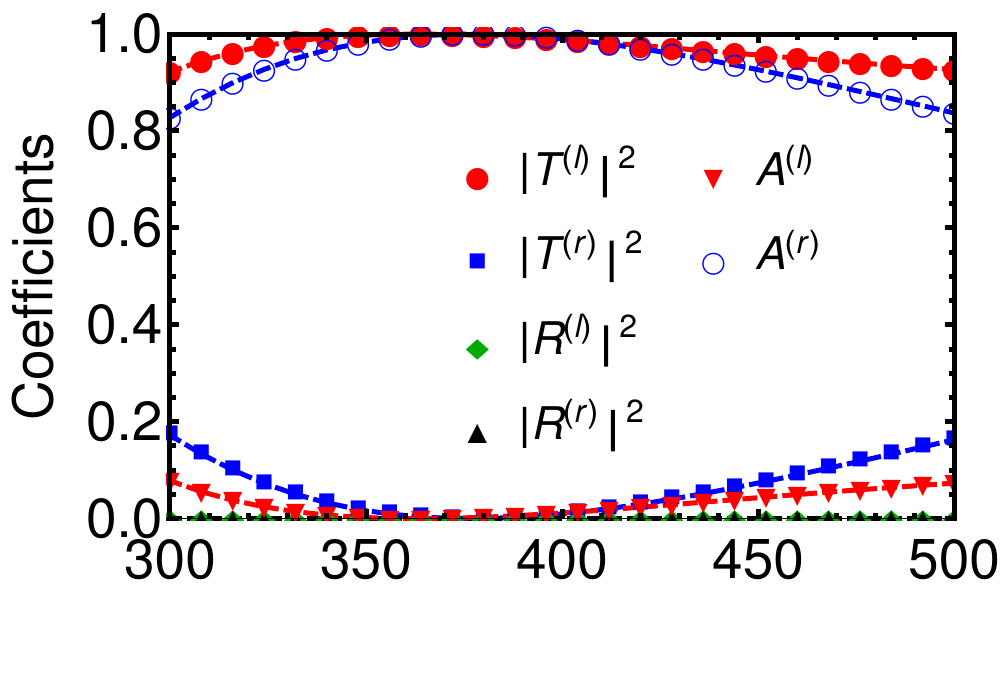}\\
\includegraphics[width=0.48\linewidth]{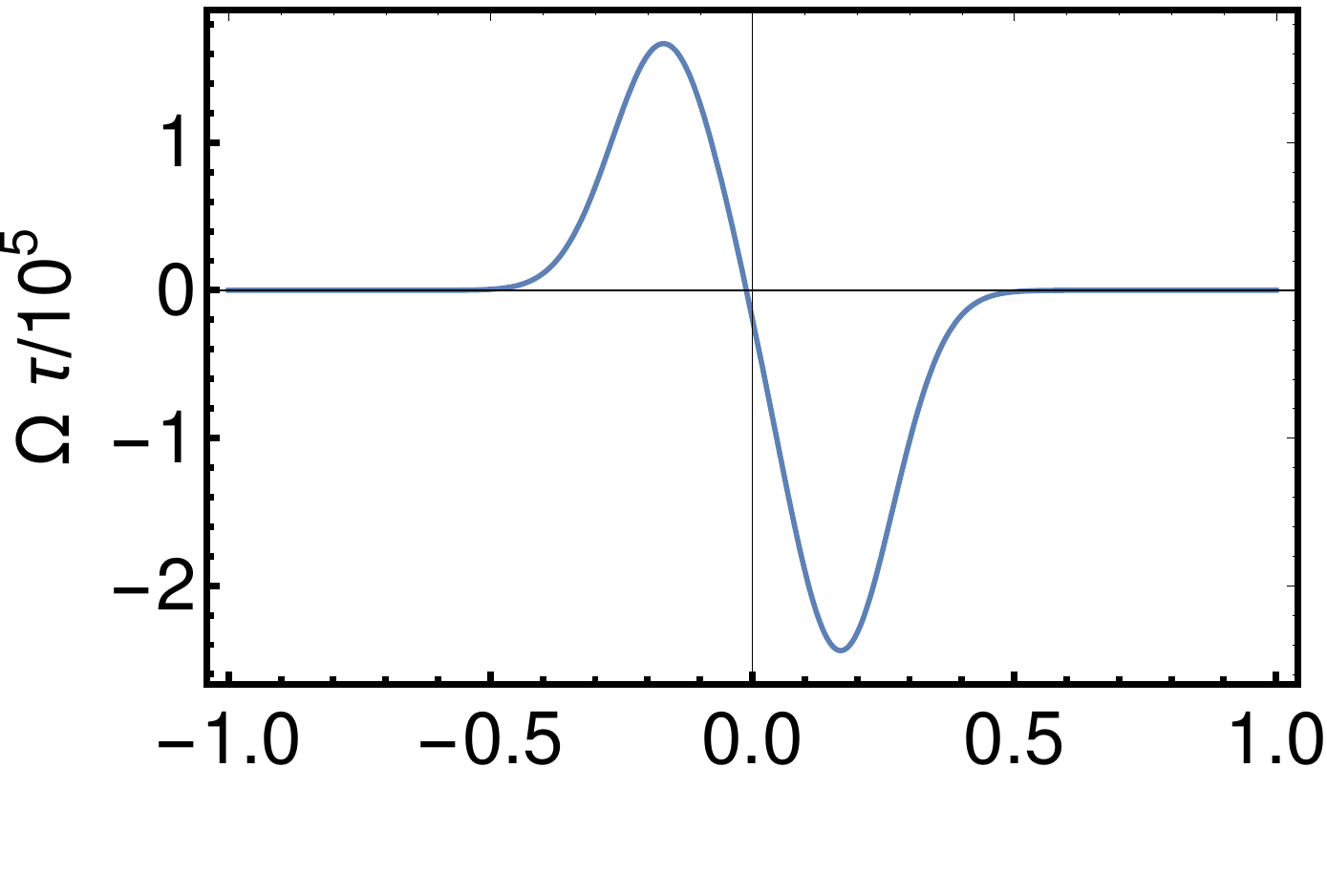}
\includegraphics[width=0.49\linewidth]{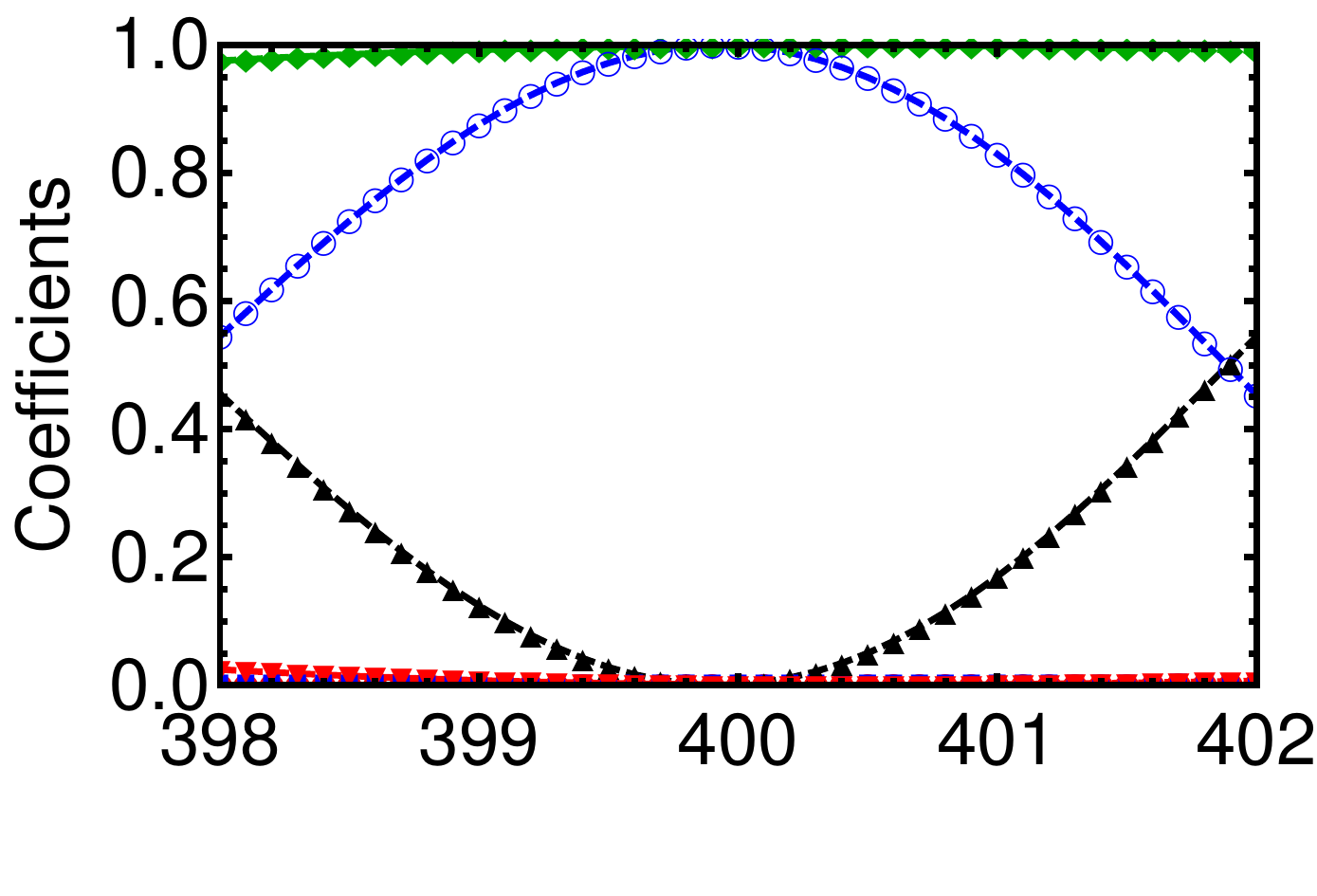}\\
\includegraphics[width=0.48\linewidth]{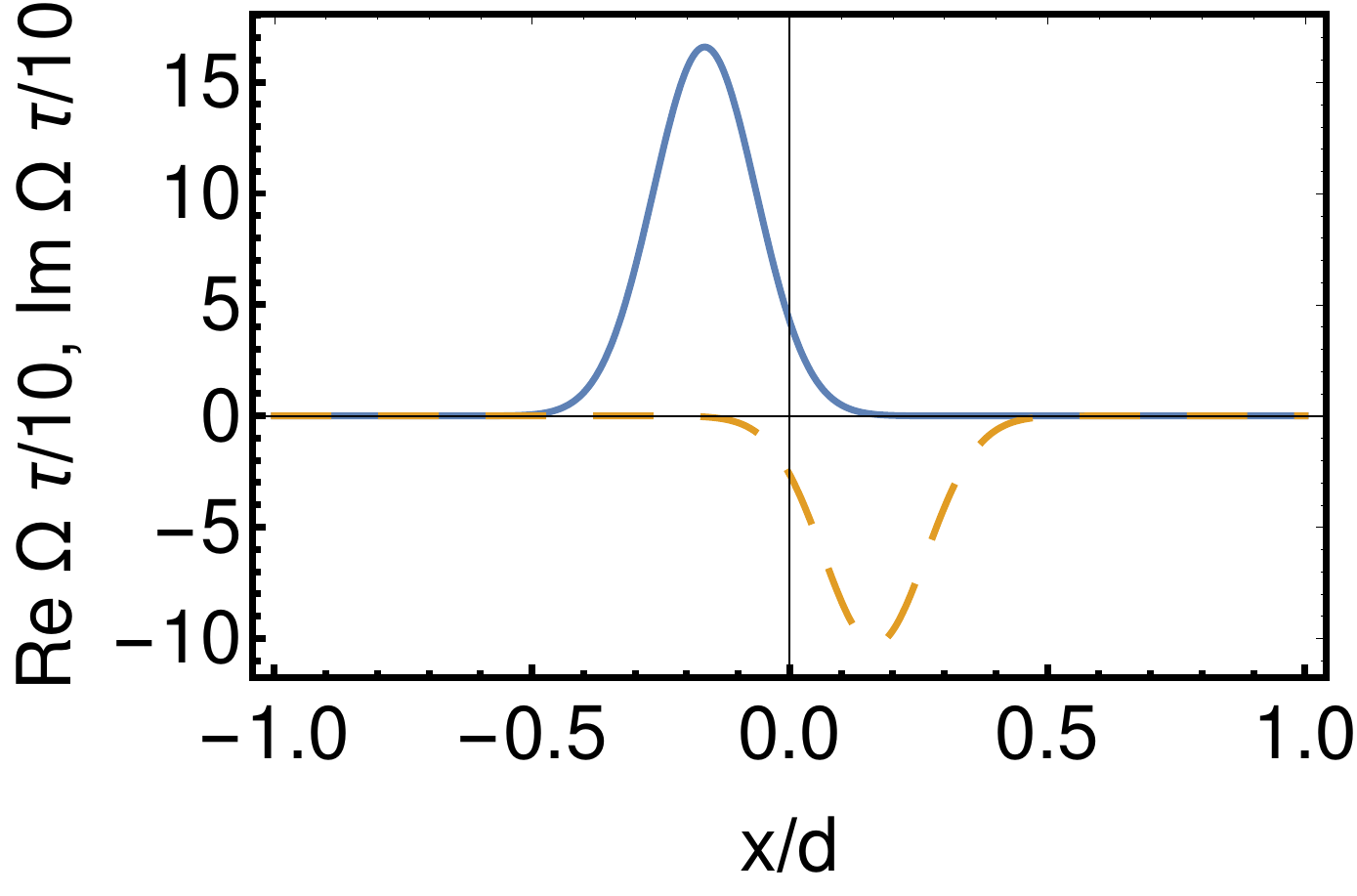}
\includegraphics[width=0.49\linewidth]{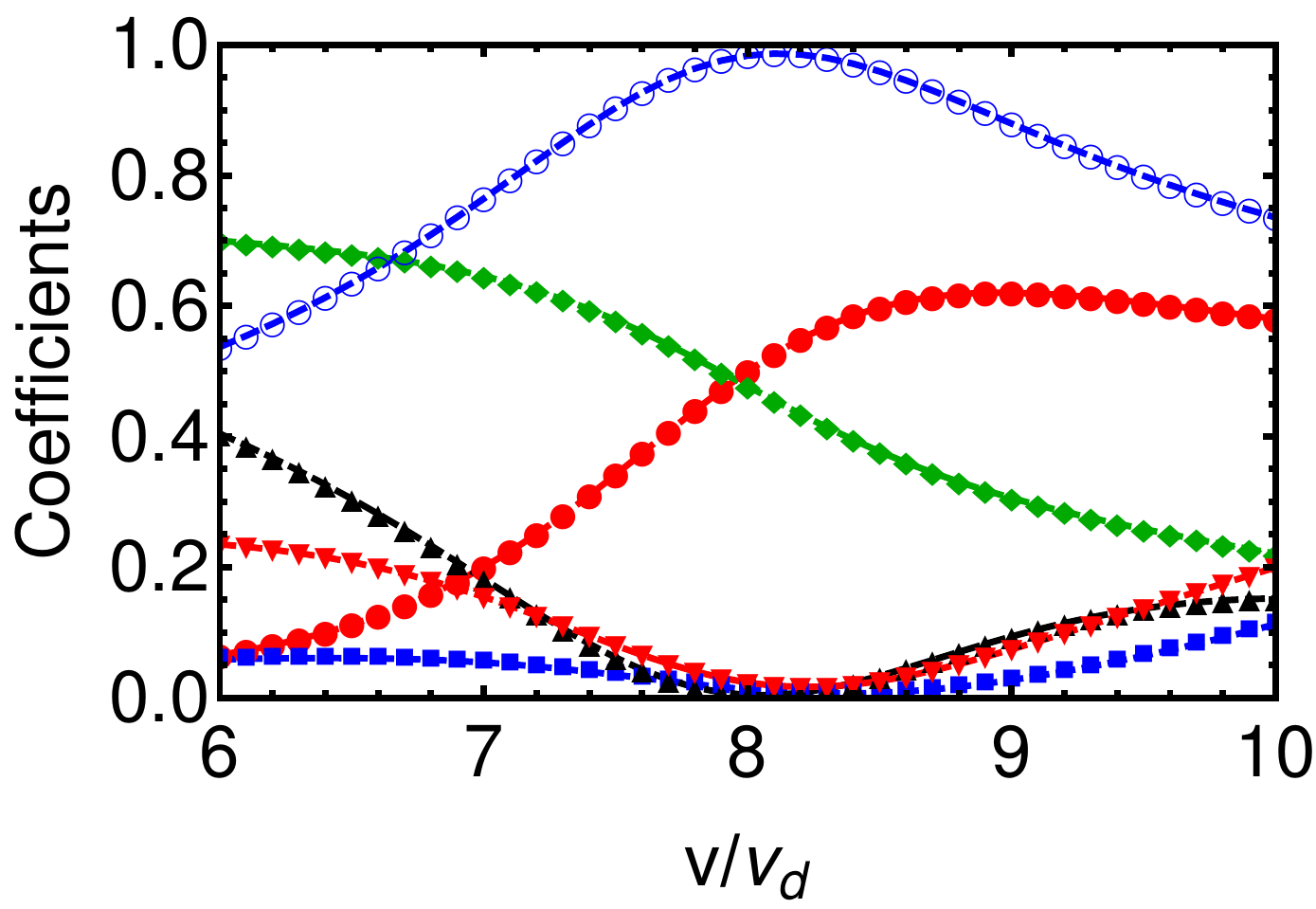}
\end{center}
\caption{Top line: ${\cal T/A}$ device with symmetry VIII.
Left: $\Omega_{\rm VIII}(x)$; 
Right:  transmission and reflection coefficients. $v_0/v_d=400$, $a\tau = 2618.19$,
$x_0/d = 0.1532$, $\tau\Delta = 1413.01$.
Middle line: ${\cal R/A}$ device with symmetry VI.
Left: $\Omega_{\rm VI} (x)$ (it is real); Right:  transmission and reflection coefficients. $v_0/v_d=400$,
$b \tau =  -244516.1$,
$c\tau = 167853.9$,
$x_0/d = 0.1679$,
$\tau\Delta= 193.508$.
Bottom line: ``Half''-${\cal TR/A}$ device with symmetry I.
Left:  $\Omega_{\rm I}(x)$, real (orange, dashed line), and imaginary parts (blue,solid line);
Right: transmission and reflection coefficients. $v_0/v_d=8$, $b\tau =  102.6520$,
$c \tau =  165.8355$,
$x_0/d = 0.1648$,
$\tau\Delta= 90.5337$. In all cases $\tau={m d^2}/{\hbar}$.
\label{fig_t_a}}
\end{figure}

\begin{figure}
\begin{center}
\includegraphics[width=0.48\linewidth]{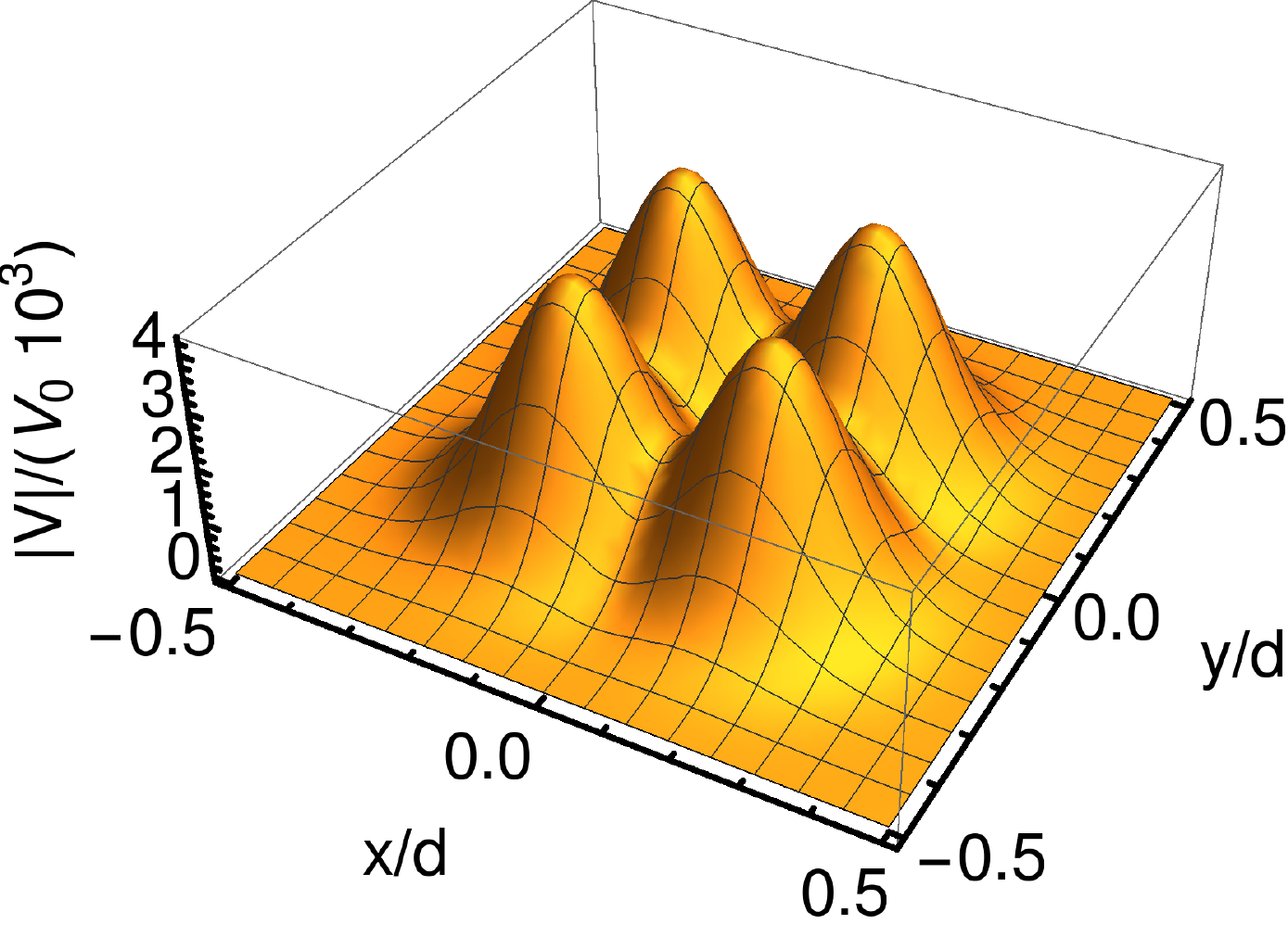}
\includegraphics[width=0.49\linewidth]{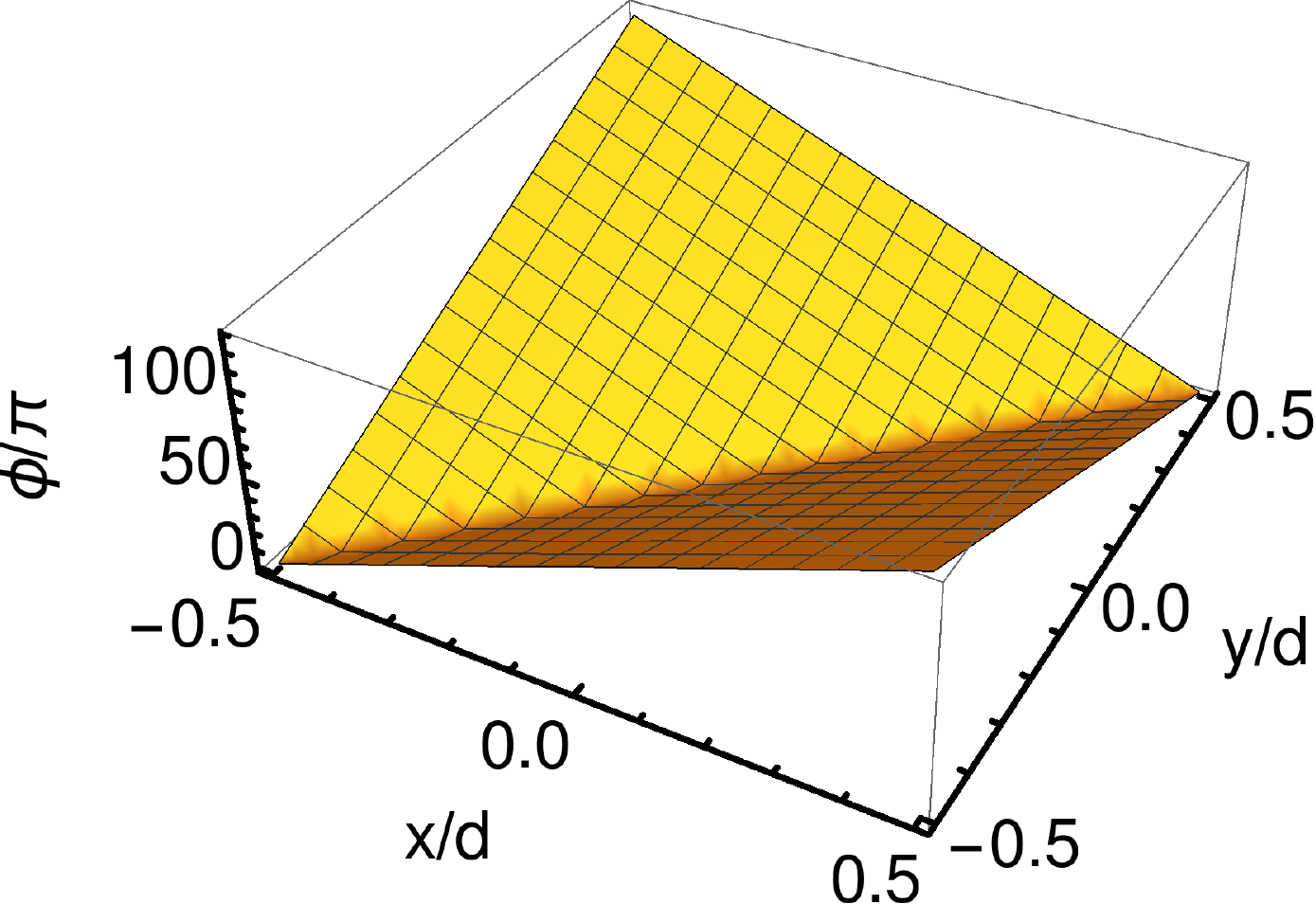}
\end{center}
\caption{Nonlocal potential $V(x,y)$ with symmetry VIII for ${\cal T/A}$ device: absolute value (left), argument (right).
$V_0=\hbar^2/(md^3)$.
\label{fig_poten1}}
\end{figure}

The three devices are worked out for $\gamma=0$, a valid approximation for  hyperfine transitions. 
We assume  for the Rabi frequencies the forms
\begin{eqnarray}
\Omega_{\rm VIII} (x) &=& a [g(x+x_0) + i  g(x-x_0)],
\nonumber\\
\Omega_{\rm VI} (x) &=&  b g(x+x_0) + c g(x-x_0),
\nonumber\\
\Omega_{\rm I} (x) &=&  - i b g(x+x_0) + c g(x-x_0),
\label{3omegas}
\end{eqnarray}
in terms of smooth, realizable Gaussians $g(x) =  \exp[-{x^2}/{w^2}]$.
We fix $2 d$ as an effective finite width 
of the potential area beyond which the potential is negligible and assumed to vanish. In the calculations we take   
$w/d=2^{1/2}/10$, 
and  set a target velocity $v_0$ to achieve the desired asymmetric scattering response. 
The real parameters $a$, $b$, $c$, $x_0$ in Eq. (\ref{3omegas}),   and  $\Delta$    
are numerically optimized with the GRAPE (Gradient Ascent Pulse Engineering) algorithm \cite{grape1,grape2}.
The Rabi frequencies will fulfill the indicated symmetries VIII, VI, and I.
The corresponding Rabi frequencies $\Omega(x)$ are  depicted in Figs. \ref{fig_t_a}, left column.
The scattering coefficients are shown in the right column. The effective non-local potential $V(x,y)$ which we are physically realising, see Eq. \eqref{effpot}, for the ${\cal T/A}$ device is shown in Fig. \ref{fig_poten1}, and the other potentials  are depicted in 
\suppB.  In the figures we use as a scaling factor for the velocity
$v_{d} = {\hbar}/({m d})$, and for time $\tau={m d^2}/{\hbar}$.  
$\Omega_{\rm VI}(x)$ should not be even (i.e. $b \neq c$) to avoid symmetry II. In addition, $\Omega_{\rm I}(x)$ should not fulfill any other symmetry than ${\rm I}$.
Fig. \ref{fig_t_a} demonstrates that the three potentials satisfy the asymmetric response conditions imposed 
at the selected velocity and also in a region nearby. 

The asymmetric behavior of the ${\cal T/A}$ device can be intuitively understood based on a classical approximation of the motion and the non-commutativity of rotations on the Bloch sphere, \seesuppC. The classical approximation also gives good estimates for the potential parameters $a$ and the detuning $\Delta$ (with a given $w$),
namely $a \approx \frac{v_0}{w} \sqrt{\pi}\, ({2}/{3})^{3/2}$ and $\Delta \approx \frac{v_0}{2w} \sqrt{\pi}\, ({2}/{3})^{3/2}$ (\seesuppC { }for details). This allows to find good initial values for further numerical optimization.
The device ${\cal T/A}$ is feasible for an experimental implementation  as the ratio $v_0/v_d$ can be 
easily increased to desired values, for  reasonable values of the
Rabi frequency and laser waist \cite{Zeyen2016} \suppB. Moreover the velocity window is quite broad. Hyperfine transitions of Beryllium ions provide an adequate two-level system for which $\gamma\approx 0$ is indeed realistic \suppB.

The ``half''-${\cal TR/A}$  device fullfills $\fabsq{T^l} = \fabsq{R^l} = 1/2$ and full absorption from the right.
The potential we use for that device has symmetry I only, i.e., ``no symmetry'' other than the trivial commutation with the identity.
No other potential symmetry would allow this type of device. The corresponding non-local potentials for the ${\cal R/A}$ and ``half''-${\cal TR/A}$  are represented
\insuppB.

{\it Discussion.}
Non-Hermitian Hamiltonians display many interesting phenomena which are 
impossible for a  Hermitian Hamiltonian  acting on the same Hilbert space. In particular, in the Hilbert space of 
a single, structureless particle on a line formed by square integrable normalizable functions, Hermitian Hamiltonians do not allow, within a linear theory \cite{Xu2014}, for asymmetric scattering transmission and reflection coefficients.
However, 
non-Hermitian Hamiltonians do.   Since devices of technological interest, such as one-way filters for transmission or reflection, one-way barriers, one-way mirrors, and others, may be built based on such scattering response asymmetries, there is both fundamental
interest and applications in sight to implement Non-Hermitian scattering Hamiltonians.    This paper is a step forward in that direction, specifically we propose a quantum-optical implementation of potentials with asymmetric scattering response. 
They are non-local and non-PT symmetrical, which allows for asymmetric transmission.      
   
In general the chosen Hilbert space may  be regarded as a subspace of a larger space. For example,  the space of a ``structureless particle'' in 1D is the ground-state subspace    
for a particle with internal structure, consisting of two-levels in the simplest scenario. 
It is then possible to regard the Non-Hermitian physics in the reduced space 
as a projection of the larger space, which may itself be driven by a  Hermitian or a Non-Hermitian Hamiltonian. 
We have seen the Hermitian option in our examples, where we assumed a zero decay constant, $\gamma=0$, for the excited state. 
A non-zero $\gamma$ implies  a Non-Hermitian  Hamiltonian in the larger two-level space. The description may still be    
enlarged,  including  quantized field modes to account for the atom-field interaction with a Hermitian Hamiltonian. 
As an outlook, depending on the application, there might be the need for a more fundamental and detailed descriptive level. Presently we discuss the desired physics (i.e., the scattering asymmetries) at the level of the smallest 1D space of the ground state, while taking refuge in the 
two-level space to find a feasible physical implementation.

\acknowledgments{We dedicate this work to R. F. Snider and C. G. Hegerfeldt for their mentorship along the years.  
This work was supported by the Basque Country Government (Grant No. IT986-16), and 
by   PGC2018-101355-B-I00 (MCIU/AEI/FEDER,UE). }

\appendix
\counterwithin{figure}{section}
\numberwithin{equation}{section}


\setcounter{equation}{0}
\setcounter{section}{1}
\section{Appendix A:  Numerical calculation of transmission and reflection coefficients}
Here we will discuss how to numerically solve  the
stationary Schr\"odinger equation for the two-level system 
by the invariant imbedding method \cite{singer.1982,band.1994}.

{\it Units.}
Let the potential ${\cal V}(x)$ be non-zero in the region $-d < x < d$. 
We  introduce the following dimensionless variables: $\bar k = (2mE)^{1/2}2d/\hbar$, $\bar x = x/(2d) + 1/2$,
$\bar\Omega (\bar x)= (4md^2/\hbar) \Omega(x)$ and $\bar\Gamma = (4md^2/\hbar) (\gamma-2i\Delta)$.
The non-Hermitian dimensionless Hamiltonian for the system takes the form
\begin{eqnarray}
\bar {\cal H}&=& \bar {\cal H}_{0}+\bar {\cal V}(\bar x), \\
\bar {\cal H}_{0}&=&- \frac{\partial^2 }{\partial {\bar x}^2}+\left(\begin{array}{cc}
0 & 0 \\
0 & -i\bar\Gamma
\end{array}\right), \\
\bar {\cal V}(\bar x)&=& \left(\begin{array}{cc}
0 & \bar\Omega (\bar x)\\
\bar\Omega (\bar x)^{*} & 0 
\end{array}\right). 
\end{eqnarray}
To set the matrices we use as in the main text the convention for internal states $\ket{1} = \left(\begin{smallmatrix}1\\0\end{smallmatrix}\right)$ and $\ket{2} = \left(\begin{smallmatrix}0\\1\end{smallmatrix}\right)$. 
To simplify the notation, we will from now on drop the bars above variables and operators for the remaining part of this section A.
The corresponding stationary Schr\"odinger equation is now
\begin{eqnarray}
k^{2} \psi^{(1)}(x)&=&-\frac{\partial^2}{\partial x^2}\psi^{(1)}(x)+\Omega(x) \psi^{(2)}(x), 
\nonumber \\
k^{2} \psi^{(2)}(x)&=&-\frac{\partial^2}{\partial x^2}\psi^{(2)}(x)+\Omega(x)^* \psi_{1}(x)-i \Gamma \psi^{(2)}(x).
\nonumber
\end{eqnarray}
Let us denote as  $|{\Psi}_\alpha(x)\ra$  the wave vector for the atom impinging in internal level $\alpha$, $\alpha=1,2$.
This vector has ground and excited state components, generically $\braket{\beta}{\psi_{\alpha}(x)}$, $\beta=1,2$, which are still functions of $x$.  
We can define the matrices $F(x)$ and $\widetilde{F}(x)$ as
\begin{eqnarray}
F_{\beta,\alpha} (x) = \braket{\beta}{\psi_{\alpha}(x)}, 
\quad
\widetilde{F}_{\beta,\alpha} (x) = \braket{\beta}{\widetilde{\psi}_{\alpha}(x)},
\end{eqnarray}
so the stationary Schr\"odinger equation can be rewritten as 
\begin{eqnarray}
\left[k^2-{\cal H}_{0}-{\cal V}(x)\right]F(x)&=&0, 
\nonumber\\
\left[k^2-{\cal H}_{0}-{\cal V}(x)\right]\widetilde{F}(x)&=&0.
\end{eqnarray}

{\it Free motion, ${\cal V}=0$.}
When ${\cal V}(x)=0$ we get
\begin{eqnarray}
\left[k^2-{\cal H}_{0}\right]\ket{\psi_{\alpha}(x)}&=&0, 
\nonumber\\
\left[k^2-{\cal H}_{0}\right]\ket{\widetilde{\psi}_{\alpha}(x)}&=&0,
\end{eqnarray}
for $\alpha=1,2$.
We can write down the solutions for particles ``coming'' from the left $\ket{\psi_{\alpha}(x)}$ in  internal state $\ket{\alpha}$ as
\begin{eqnarray}
\ket{\psi_{1}(x)} = \left(\begin{array}{c}
\frac{1}{\sqrt{k}} e^{i k x}\\
0
\end{array}\right)\!,
\, 
\ket{\psi_{2}(x)} = \left(\!\!\begin{array}{c}
0\\
\frac{1}{\sqrt[4]{k^2+i \Gamma}} e^{i \sqrt{k^2+i \Gamma}x} 
\end{array}\!\!\right)\!, 
\nonumber
\end{eqnarray}
where we assume the branch $\operatorname{Im} \sqrt{k^2+i \Gamma}\ge 0$.
$\ket{\psi_2(x)}$ is a regular traveling wave only for real $\sqrt{k^2+i\Gamma})$. If the square root has an imaginary part,  $\ket{\psi_2(x)}$  decays  from left to right.   
The solutions for incidence from the right  $\ket{\widetilde{\psi}_{\alpha}(x)}$ in internal state $\ket{\alpha}$ are similarly
\begin{eqnarray}
\ket{\widetilde{\psi}_{1}(x)} = \left(\!\begin{array}{c}
\frac{1}{\sqrt{k}} e^{-i k x}\\
0 
\end{array}\!\right)\!, 
\ket{\widetilde{\psi}_{2}(x)} = \left(\!\!\begin{array}{c}
0\\
\frac{1}{\sqrt[4]{k^2+i \Gamma}} e^{-i \sqrt{k^2+i \Gamma}x} 
\end{array}\!\!\right)\!. 
\nonumber
\end{eqnarray}
The normalization is chosen in such a way that the dimensionless probability current
is constant (and equal) for all solutions with real $\sqrt{k^2+i\Gamma}$. 

The solutions are given by $F(x) = h_+ (x)$ and $\widetilde F (x) = h_- (x)$, where
\begin{equation}
h_{\pm}(x)=\left(\begin{array}{cc}
\frac{1}{\sqrt{k}}e^{\pm i k x} & 0\\
0 & \frac{1}{\sqrt[4]{k^2+i \Gamma}} e^{\pm i \sqrt{k^2+i \Gamma} x} \\
\end{array}\right).
\end{equation}
The Wronskian is $W(h_{+},h_{-})(x)=2i$ so that these are linearly independent solutions.

{\it General case.}
To solve the general case, we construct the Green's function defined by
\begin{equation}
(k^2-{\cal H}_{0})G_{0}(x,x')=\delta(x-x')\mathbf{1}.
\end{equation}
It is given by 
\begin{eqnarray}
G_{0}(x,x')&=&W^{-1} \begin{cases} 
h_{+}(x)h_{-}(x')  & x>x' , \\ h_{+}(x')h_{-}(x) & x'>x ,
   \end{cases} \\
   &=& -\frac{i}{2} \left(\!\begin{array}{cc}
\frac{1}{k}e^{ i k \fabs{x-x'}} & 0\\
0 & \frac{e^{ i \sqrt{k^2+i \Gamma} \fabs{x-x'}}}{\sqrt{k^2+i \Gamma}} \\
\end{array}\!\right). \nonumber
\end{eqnarray}
The Green's function allows us to solve for   $F(x)$ and $\widetilde{F}(x)$ in integral form,
\begin{eqnarray}
F(x)=h_{+}(x)+\int_{-\infty}^{\infty} dx' G_{0}(x,x') {\cal V}(x') F(x'), 
\nonumber\\
\widetilde{F}(x)=h_{-}(x)+\int_{-\infty}^{\infty} dx' G_{0}(x,x') {\cal V}(x') \widetilde{F}(x').
\label{eqf}
\end{eqnarray}

{\it Asymptotic form of the solutions.}
From Eq. (\ref{eqf}) we find the following asymptotic forms of  $F(x)$ and $\widetilde{F}(x)$:
\begin{eqnarray}
F_{\eta}(x)&=& \begin{cases} 
h_{+}(x)+h_{-}(x)R  & x<0  \\ h_{+}(x)T & x>1 
   \end{cases},
\nonumber\\   
\widetilde{F}_{\eta}(x)&=&\begin{cases} 
h_{-}(x)\widetilde{T}  &  x<0  \\ h_{-}(x)+h_{+}(x)\widetilde{R} & x>1 
   \end{cases},
\end{eqnarray}
where the $R$ and $T$ matrices for incidence from the left are given by
\begin{eqnarray}
R &=& W^{-1}\int_{0}^{1} dx' h_{+}(x'){\cal V}(x')F (x'), 
\nonumber\\
T &=& \mathbf{1}+W^{-1}\int_{0}^{1} dx' h_{-}(x'){\cal V}(x')F (x'),
\end{eqnarray}
whereas, for right incidence, 
\begin{eqnarray}
\widetilde{R} &=& W^{-1}\int_{0}^{\eta} dx' h_{-}(x'){\cal V}(x')\widetilde{F}_{\eta}(x'), 
\nonumber\\
\widetilde{T} &=& \mathbf{1}+W^{-1}\int_{0}^{\eta} dx' h_{+}(x'){\cal V}(x')\widetilde{F}_{\eta}(x').
\end{eqnarray}
In particular, for left incidence  in the ground-state, we get  if $x < 0$,
\begin{eqnarray}
\hspace*{-.5cm}|\psi_1 (x)\ra = \left(\!\!\begin{array}{c} \frac{1}{\sqrt{k}}e^{i k x} \\ 0\end{array}\!\!\right) 
+ \left(\!\!\begin{array}{c} 
R_{1,1} \frac{1}{\sqrt{k}}e^{-i k x} \\
R_{2,1} \frac{1}{\sqrt[4]{k^2+i \Gamma}} e^{- i \sqrt{k^2+i \Gamma} x}
\end{array}\!\!\right)\!,
\label{as-}
\end{eqnarray}
and, if $x>1$,
\begin{eqnarray}
|\psi_1 (x)\ra = \left(\begin{array}{c} 
T_{1,1} \frac{1}{\sqrt{k}}e^{i k x} \\
T_{2,1} \frac{1}{\sqrt[4]{k^2+i \Gamma}} e^{i \sqrt{k^2+i \Gamma} x}
\end{array}\right).
\label{as+}
\end{eqnarray}
When $\sqrt{k^2+i \Gamma}$ is real, the elements of $T$ and $R$ in Eqs. (\ref{as-}) and (\ref{as+}) 
are transmission and reflection amplitudes for waves traveling away from the interaction region. 
However when  ${\rm Im}\sqrt{k^2+i \Gamma}>0$ the waves for the excited state $2$ are evanescent. 
In scattering theory parlance the channel is ``closed'', so  the $T_{2,1}$ and $R_{2,1}$ are just proportionality factors 
rather than proper transmission 
and reflection amplitudes for travelling waves. By continuity however, it is customary to keep the same notation 
and even terminology for closed or open channels.

In a similar way, for right incidence in the ground state and 
$x > 1$,
\begin{eqnarray}
\hspace*{-.5cm}|\widetilde\psi_1 (x)\ra = \left(\!\!\begin{array}{c} \frac{1}{\sqrt{k}}e^{-i k x} \\ 0\end{array}\!\!\right) 
\!+\! \left(\!\!\begin{array}{c} 
\widetilde R_{1,1} \frac{1}{\sqrt{k}}e^{i k x}
\\
\widetilde R_{2,1} \frac{1}{\sqrt[4]{k^2+i \Gamma}} e^{i \sqrt{k^2+i \Gamma} x}
\end{array}\!\!\right)\!\!,
\end{eqnarray}
whereas, for $x<0$,
\begin{eqnarray}
|\widetilde \psi_1 (x)\ra = \left(\begin{array}{c} 
\widetilde T_{1,1} \frac{1}{\sqrt{k}}e^{-i k x} \\
\widetilde T_{2,1} \frac{1}{\sqrt[4]{k^2+i \Gamma}} e^{-i \sqrt{k^2+i \Gamma} x}
\end{array}\right).
\end{eqnarray}
Note that alternative definitions of the amplitudes may be found in many works,
without momentum prefactors.   

The amplitudes relevant for the main text are $T^l=T_{1,1}$, 
$T^r=\widetilde{T}_{1,1}$, $R^l=R_{1,1}$, and $R^r=\widetilde{R}_{1,1}$.  The following 
subsection explains how to compute them.    

{\it Differential equations for $R$ and $T$ matrices.}
To solve for $R$ and $T$ we will use cut-off versions of the  potential,
\begin{equation}
{\cal V}_{\eta}=\begin{cases} 
{\cal V}(x)  & 0\leq\eta\leq 1 , \\ 0& \text{Otherwise} 
   \end{cases},
\end{equation}
where $0 \le \eta \le 1$, and corresponding matrices 
\begin{eqnarray}
R_{\eta}&=& W^{-1}\int_{0}^{\eta} dx' h_{+}(x'){\cal V}(x')F_{\eta}(x'), 
\nonumber\\
T_{\eta}&=& \mathbf{1}+W^{-1}\int_{0}^{\eta} dx' h_{-}(x'){\cal V}(x')F_{\eta}(x'),
\nonumber\\
\widetilde{R}_{\eta} &=& W^{-1}\int_{0}^{\eta} dx' h_{-}(x'){\cal V}(x')\widetilde{F}_{\eta}(x'), 
\nonumber\\
\widetilde{T}_{\eta} &=& \mathbf{1}+W^{-1}\int_{0}^{\eta} dx' h_{+}(x'){\cal V}(x')\widetilde{F}_{\eta}(x').
\end{eqnarray}
Taking the derivative of these matrices with respect to $\eta$, we find a set
of coupled differential equations,
\begin{eqnarray}
\frac{d R_{\eta}}{d \eta}&=& W^{-1} \widetilde{T}_{\eta} h_{+}(\eta){\cal V}(\eta)h_{+}(\eta)T_{\eta}, 
\nonumber\\
\frac{d T_{\eta}}{d \eta}&=& W^{-1} \left[h_{-}(\eta)+\widetilde{R}_{\eta}h_{+}(\eta)\right]{\cal V}(\eta)h_{+}(\eta)T_{\eta}, 
\nonumber\\
\frac{d \widetilde{R}_{\eta}}{d \eta}&=& W^{-1}\!\!\left[h_{-}(\eta)\!+\!\widetilde{R}_{\eta}h_{+}(\eta)\right]\!{\cal V}(\eta)\!\left[h_{-}(\eta)\!+\!h_{+}(\eta)\widetilde{R}_{\eta}\right]\!, 
\nonumber\\
\frac{d \widetilde{T}_{\eta}}{d \eta}&=& W^{-1} \widetilde{T}_{\eta} h_{+}(\eta){\cal V}(\eta)\left[h_{-}(\eta)+h_{+}(\eta)\widetilde{R}_{\eta}\right].
\label{a20}
\end{eqnarray}
The initial conditions are $R_{0} = \widetilde{R}_{0}=0$ and $T_{0} = \widetilde{T}_{0}=\mathbf{1}$.

{\it Improving numerical efficiency.}
The last two equations involve only  matrices for incidence from the right, they  do not couple to any left-incidence matrix, whereas the equations for left incidence 
amplitudes involve couplings with amplitudes for right incidence. This asymmetry is due to the way we do the potential slicing. The asymmetry  is not ``fundamental'' 
but we can use it for our advantage to simplify calculations. We can solve the two last  equations to get amplitudes for right incidence. 
To get amplitudes for left incidence we use a mirror image of the potential and solve also the last two equations.   
Thus it is enough to find an efficient numerical method to solve the last two equations.
In principle, one can now solve these differential equations from $\eta=0$ to $1$ to get all reflection and transmission amplitudes using the boundary conditions $\widetilde{R}_{0}=0$ and $\widetilde{T}_{0}=\mathbf{1}$. However due to the exponential nature of the 
free-space solutions $h_{\pm}(x)$ especially if ${\rm Im}\sqrt{k^2+i \Gamma}>0$, this is not very efficient numerically.

To avoid this problem we make new definitions,
\begin{eqnarray}
\hat{S}_{\eta}&=&\mathbf{1}+h_{+}(\eta)\widetilde{R}_{\eta}h_{-}^{-1}(\eta), 
\nonumber\\
\hat{T}_{\eta}&=&h_{+}(0)\widetilde{T}_{\eta}h_{-}^{-1}(\eta), 
\nonumber\\
\hat{\cal V}(\eta)&=&W^{-1}h_{+}^{2}(0){\cal V}(\eta), 
\nonumber\\
\hat{Q}&=&i h_{+}^{-2}(0).
\end{eqnarray}
Rewriting the last two equations in Eq. (\ref{a20}) in terms of these new variables we get
\begin{eqnarray}
\frac{d \hat{S}_{\eta}}{d \eta}&=&-2 \hat{Q}+\hat{Q}\hat{S}_{\eta}+\hat{S}_{\eta}\left[\hat{Q}+\hat{\cal V}(\eta)\hat{S}_{\eta}\right], 
\nonumber\\
\frac{d \hat{T}_{\eta}}{d \eta}&=& \hat{T}_{\eta}\left[\hat{Q}+\hat{\cal V}(\eta)\hat{S}_{\eta}\right],
\end{eqnarray}
with initial conditions $\hat{T}_{0} =\hat{S}_{0}=\mathbf{1}$.

Let us consider solely incidence in the ground state. For right incidence in the ground state,
the reflection coefficients and transmission coefficient are
\begin{eqnarray}
\widetilde R_{1,1} &=& e^{-2ik} \left[(\hat S_{\eta=1})_{1,1} - 1 \right],
\nonumber\\
\widetilde R_{2,1} &=& 
\frac{\sqrt[4]{k^2+i \Gamma}}{\sqrt{k}} e^{-ik-i\sqrt{k^2+i \Gamma}} (\hat S_{\eta=1})_{2,1},
\nonumber\\
\widetilde T_{1,1} &=& e^{-i k} (\hat T_{\eta=1})_{1,1},
\nonumber\\ 
\widetilde T_{2,1} &=& \frac{\sqrt[4]{k^2+i \Gamma}}{\sqrt{k}} e^{-i k} (\hat T_{\eta=1})_{2,1}.
\end{eqnarray}

{\it Bounds from unitarity.}
The $S$-matrix 
\beq
S=\left(\begin{array}{cccc}
T_{11}&T_{12}&\widetilde R_{11}&\widetilde R_{12}
\\
T_{21}&T_{22}&\widetilde R_{21}&\widetilde R_{22}
\\ 
R_{11}&R_{12}&\widetilde T_{11}&\widetilde T_{12}
\\
R_{21}&R_{22}&\widetilde T_{21}&\widetilde T_{22}
\end{array}\right)
\eeq
is unitary for Hermitian Hamiltonians,  in particular when $\gamma=0$. 
Unitarity implies relations among the matrix elements and in particular 
\beqa
1&\ge& |R_{11} |^2+|T_{11} |^2,
\\
1&\ge& |\widetilde R_{11}|^2+|\widetilde T_{11}|^2,
\\
1&\ge& |\widetilde R_{11} |^2+|T_{11} |^2,
\\
1&\ge& |R_{11} |^2+|\widetilde T_{11}|^2.
\eeqa
While the first two are rather obvious because of  probability conservation, the last two are less so, and 
set physical  limits to the possible asymmetric devices that can be constructed in the ground state subspace.  
%
\setcounter{equation}{0}
\setcounter{figure}{0}
\setcounter{section}{2}
\section{Appendix B:  Explicit forms of the non-local potentials. Feasibility}
%
%

\begin{figure}
\begin{center}
\includegraphics[width=0.48\linewidth]{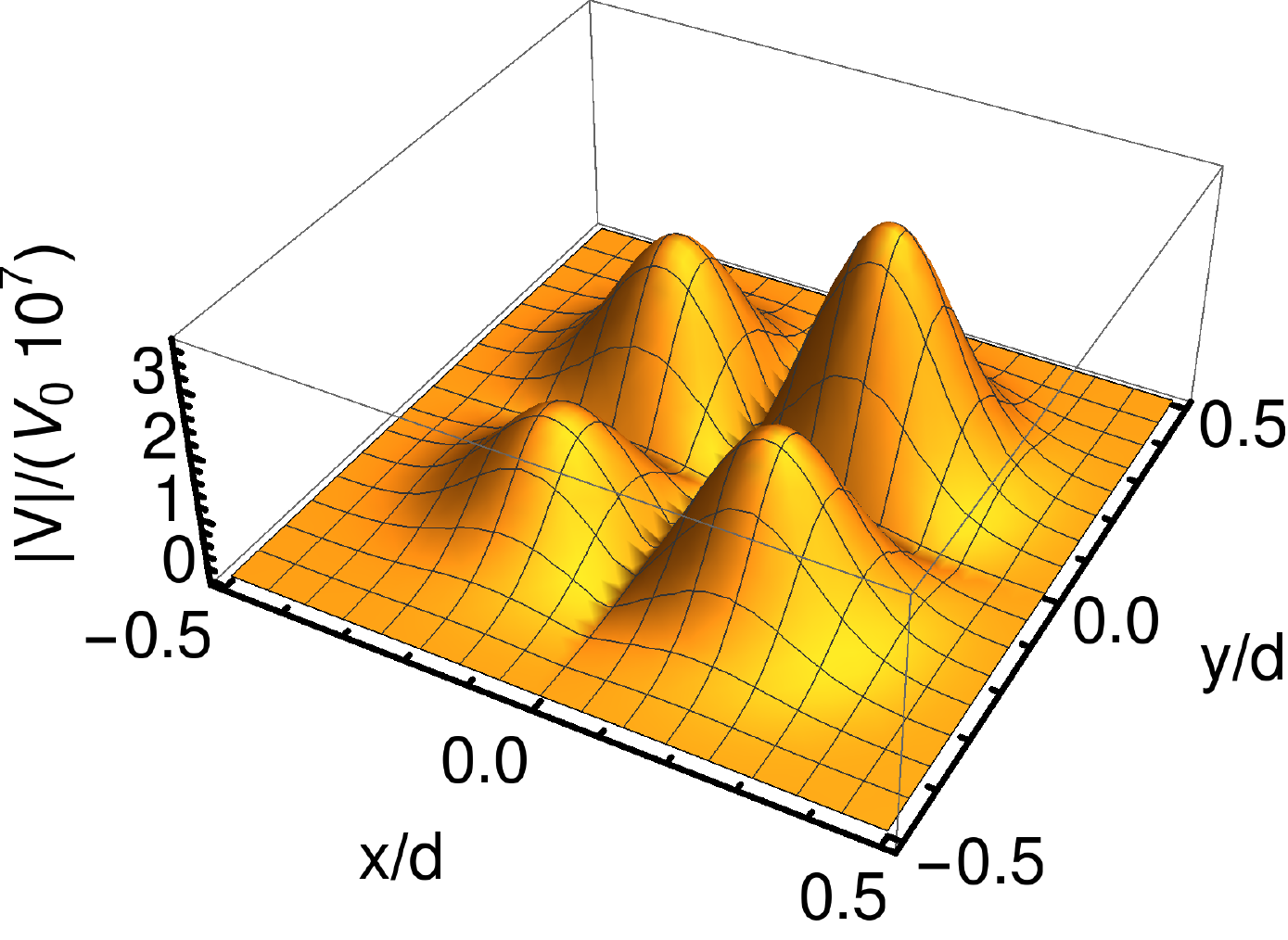}
\includegraphics[width=0.49\linewidth]{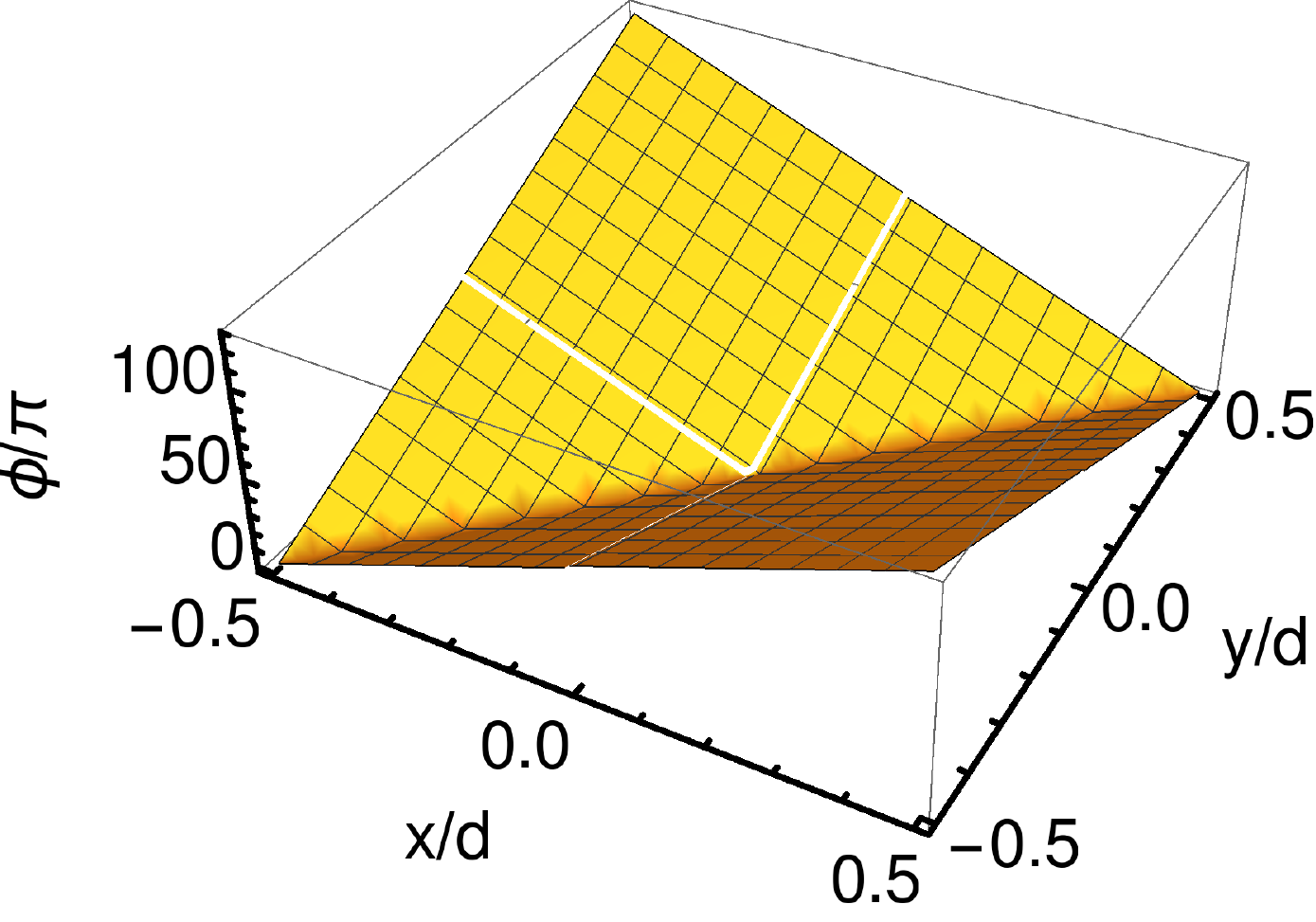}\\
\includegraphics[width=0.48\linewidth]{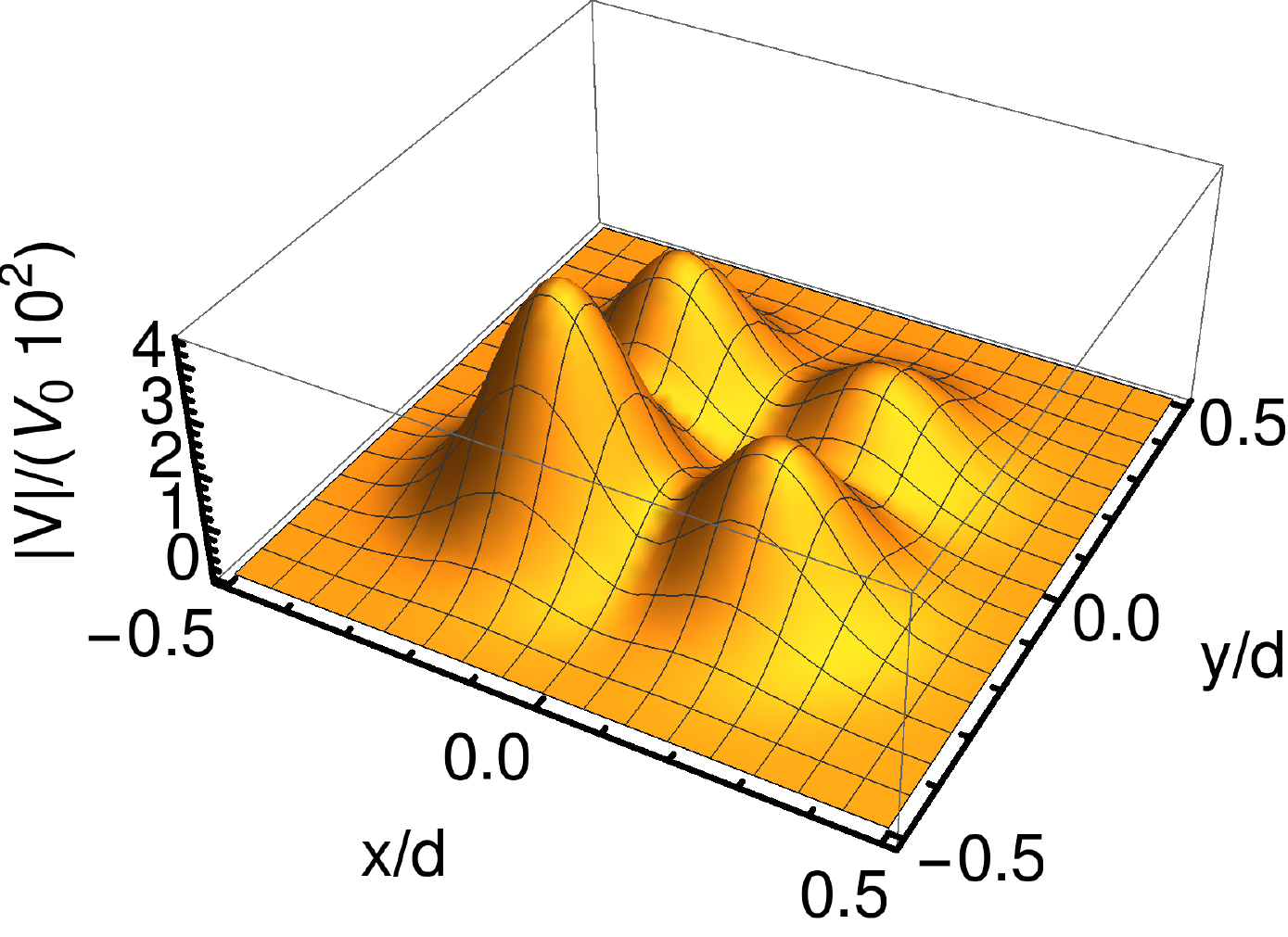}
\includegraphics[width=0.49\linewidth]{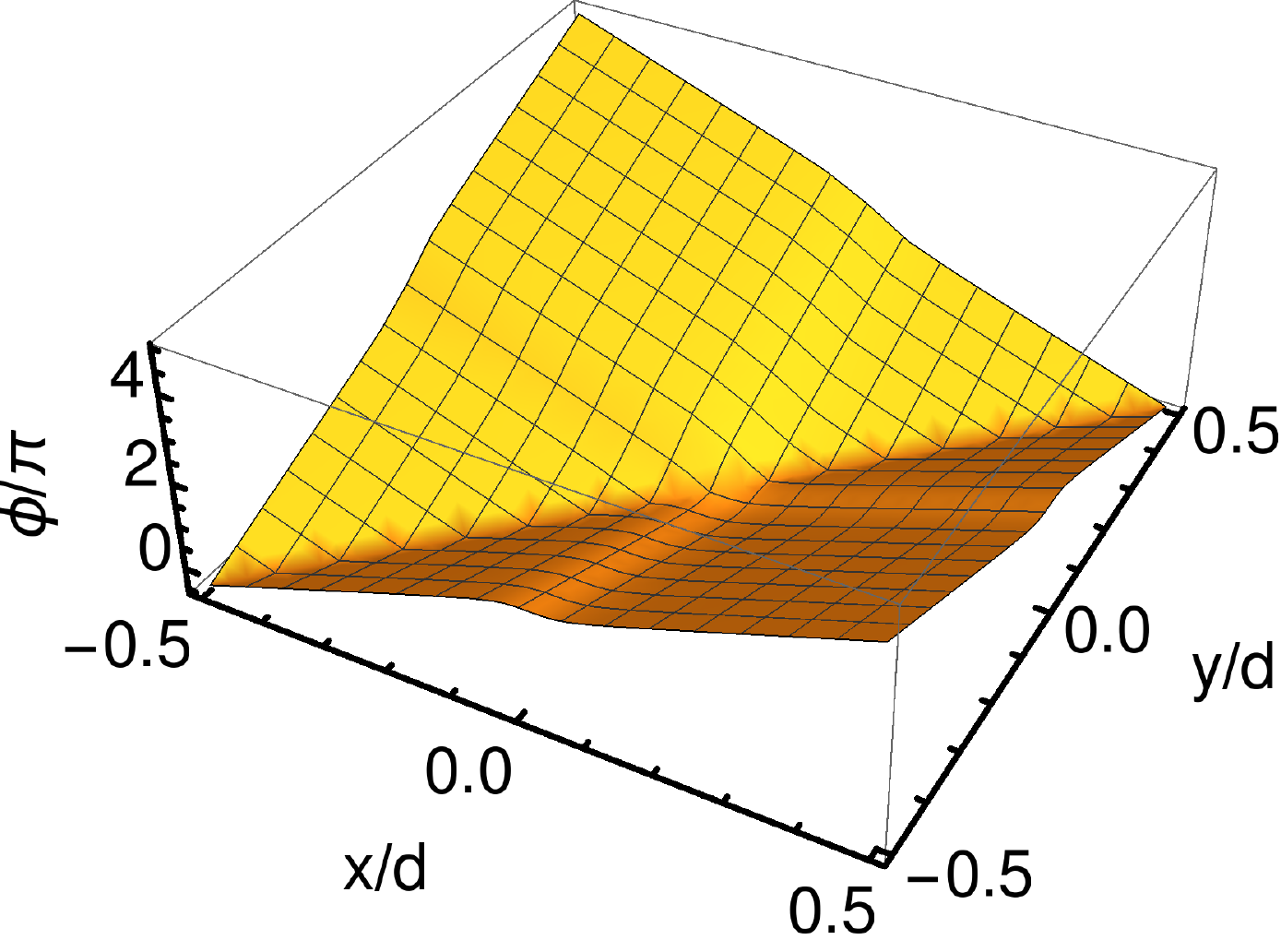}
\end{center}
\caption{Nonlocal potentials (absolute value and argument) corresponding to Fig. 1 in the main text.  
Top line: Potential for ${\cal R/A}$ device with symmetry VI.
Bottom line: ``Half''-${\cal TR/A}$ device with symmetry I.
$V_0=\hbar^2/(md^3)$. The potential for the ${\cal T/A}$ device is in Fig. 2
of the main text. 
\label{fig_poten}}
\end{figure}

Figure 2 of the main text and Fig. \ref{fig_poten} give the non-local potentials corresponding to the $v/v_d$ ratios used in Fig. 1 of the main text. 
Absolute value  and argument are provided. Note that the non-local potential has dimensions  energy/length, so  we 
divide the absolute value by a factor $V_0=\hbar^2/(m d^3)$ to plot a dimensionless quantity.  

In the parameter optimization we see that increasing the velocities further does not pose a problem for the ${\cal T/A}$ 
device, it is more challenging for a ${\cal R/A}$ device, and it is quite difficult for the half-${\cal RT/A}$ device. 
Moreover the velocity width with the desired behavior is much broader for ${\cal T/A}$. Therefore a ${\cal T/A}$ 
device is the best candidate for 
an experimental implementation.  
As a check of feasibility, let us assume a Beryllium ion. Its hyperfine structure provides a good  two-level system
for which we can neglect decay. We have $m=1.49\times 10^{-26}$ kg
and set a length $d=10\, \mu$m compatible with the small laser waists (in this case 1.4 $\mu$m) achieved for individual ion
addressing \cite{Zeyen2016}. The scaling factors take the values  
\beqa
v_d&=&0.67\, {\rm mm/s},
\nonumber\\
\tau&=&1.49\times 10^{-2}\, {\rm s},
\nonumber
\eeqa
which gives  $v\approx$ 27 cm/s for $v/v_d=400$, (again, we see no major obstacle to get devices for higher velocities,
in particular the approximations in Appendix  C can be used to  estimate the values of the parameters)
and Rabi frequencies, see Fig. 1 in main text,  in the hundreds of kHz range. The relative ion-laser beam velocity could be as well 
implemented  by moving the beam in the laboratory frame. 
\setcounter{section}{3}
\setcounter{figure}{0}
\section{Appendix C: Why is the scattering asymmetric? Intuitive answers from approximate dynamics}
\begin{figure}
\begin{center}
\includegraphics[width=0.48\linewidth]{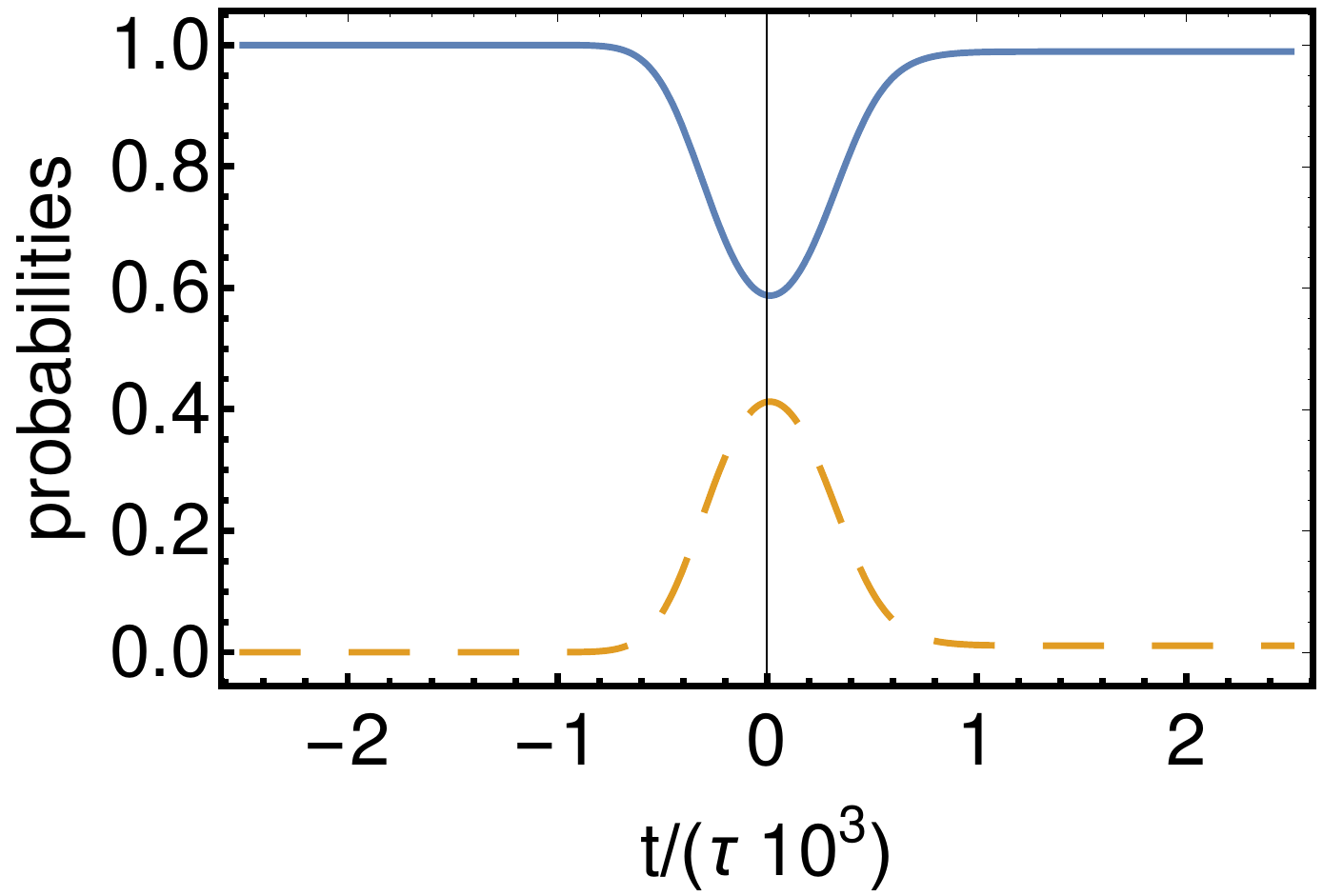}
\includegraphics[width=0.48\linewidth]{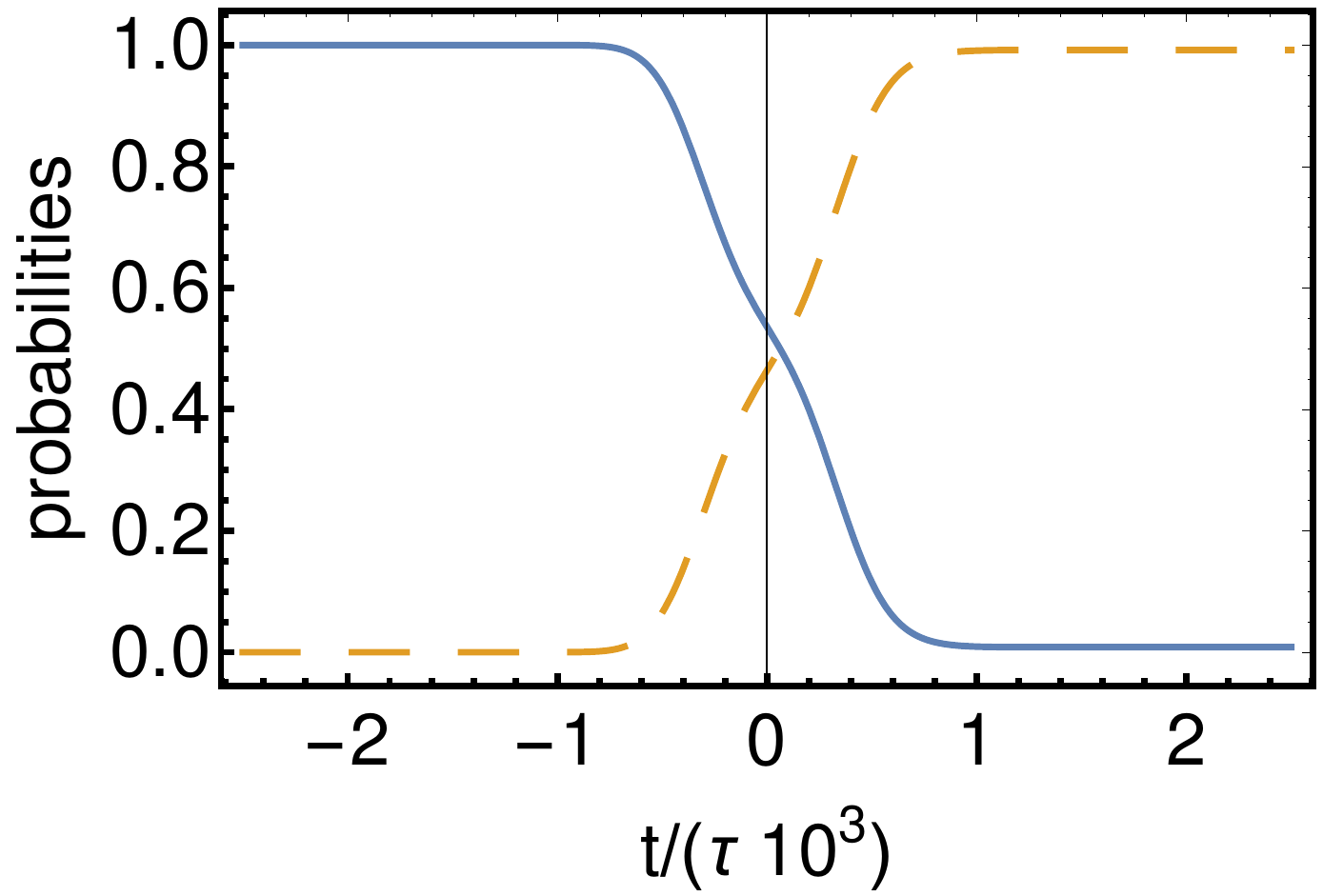}
\end{center}
\caption{Simplified model of the asymmetric ${\cal T/A}$ device with symmetry VIII: (a) $\chi_+(t)$, (b) $\chi_-(t)$; ground-state population $\fabsq{\chi_{\pm(t),1}}$ (blue, solid line), excited-
$\fabsq{\chi_{\pm(t),2}}$ (orange, dashed line). $v/v_d = 400$, $a\tau = 2618.19$,
$x_0/d = 0.1532$, $\tau\Delta = 1413.01$.
\label{fig_t_a_approx}}
\end{figure}

In a ${\cal T/A}$ device such as the one worked out in the main text an incident plane wave from the left ends up as a pure transmitted wave with no reflection or absorption. 
However, a wave incident from the right is fully absorbed. How can that be? Should not the velocity-reversed motion 
of the transmitted wave lead to the reversed incident wave?
For a more intuitive understanding we may seek help in the underlying two-level model. 
In the larger space the potential is again local and Hermitian. A simple semiclassical 
approximation is to assume that the particle moves with  constant speeds $\pm v$ for left ($v>0$) or right ($-v<0$) incidence,  so that at a given time it is subjected to  the $2\times2$ time-dependent potentials  
${\cal V}(\pm vt)$. The incidence from the left and right give different time dependences for the potential. The scattering problem then reduces to solving the time-dependent Schr\"odinger equation for the amplitudes of a two-level atom with time-dependent potential, i.e. to solving the following time-dependent Schr\"odinger equation ($\gamma = 0$)
\begin{eqnarray}
i \hbar \frac{\partial}{\partial t} \chi_\pm(t)
= {\cal V} (\pm v t) \chi_\pm(t),
\end{eqnarray}
with the appropriate boundary conditions $\chi_+ (-\infty) = \chi_- (-\infty) =\left(\begin{smallmatrix} 1\\ 0\end{smallmatrix}\right)$. The  solutions for $v/v_d = 400$ 
are shown in Fig. \ref{fig_t_a_approx}.
In Fig. \ref{fig_t_a_approx}(a), $\chi_+ (t)$ (left incidence) is depicted:  the particle ends  with high probability in the ground state at final time. In Fig. \ref{fig_t_a_approx}(b), $\chi_- (t)$ (right incidence) demonstrates  the ground state population is transferred to the excited state. Projected onto the ground-state level alone,
this corresponds to full absorption of the ground state population at final time.

For an  even rougher but also illustrative picture,  again in a semiclassical time-dependent framework, we  may substitute the smooth Gaussians for Re$(\Omega)$ and Im$(\Omega)$ in Fig. \ref{fig_t_a} by two simple, contiguous square functions of height 
$\Omega>0$ and width $\tilde{w} > 0$. Then, the $2\times2$ potential at a given time is, in terms of Pauli matrices,
\begin{eqnarray}
{\cal V} (x) = \frac{\hbar}{2}\Delta (\sigma_Z-{\mathbf 1})+ \frac{\hbar}{2} \left\{\begin{array}{cc}
\Omega\sigma_X & -\tilde w < x < 0\\
-\Omega\sigma_Y & 0 < x < \tilde w\\
0 & \mbox{otherwise}
\end{array}\right.
\end{eqnarray}
where $x = \pm v t$ and let ${\sf T}=2 \tilde w/v$.

\begin{figure}
\fbox{
\begin{minipage}{8cm}
\flushleft  (a) Order of rotations:  first  $R_1({\sf T}/2)$ (left figure) and then $R_2({\sf T}/2)$ (right figure)
\begin{center}
\vspace*{-0.32cm}
\includegraphics[width=0.49\linewidth]{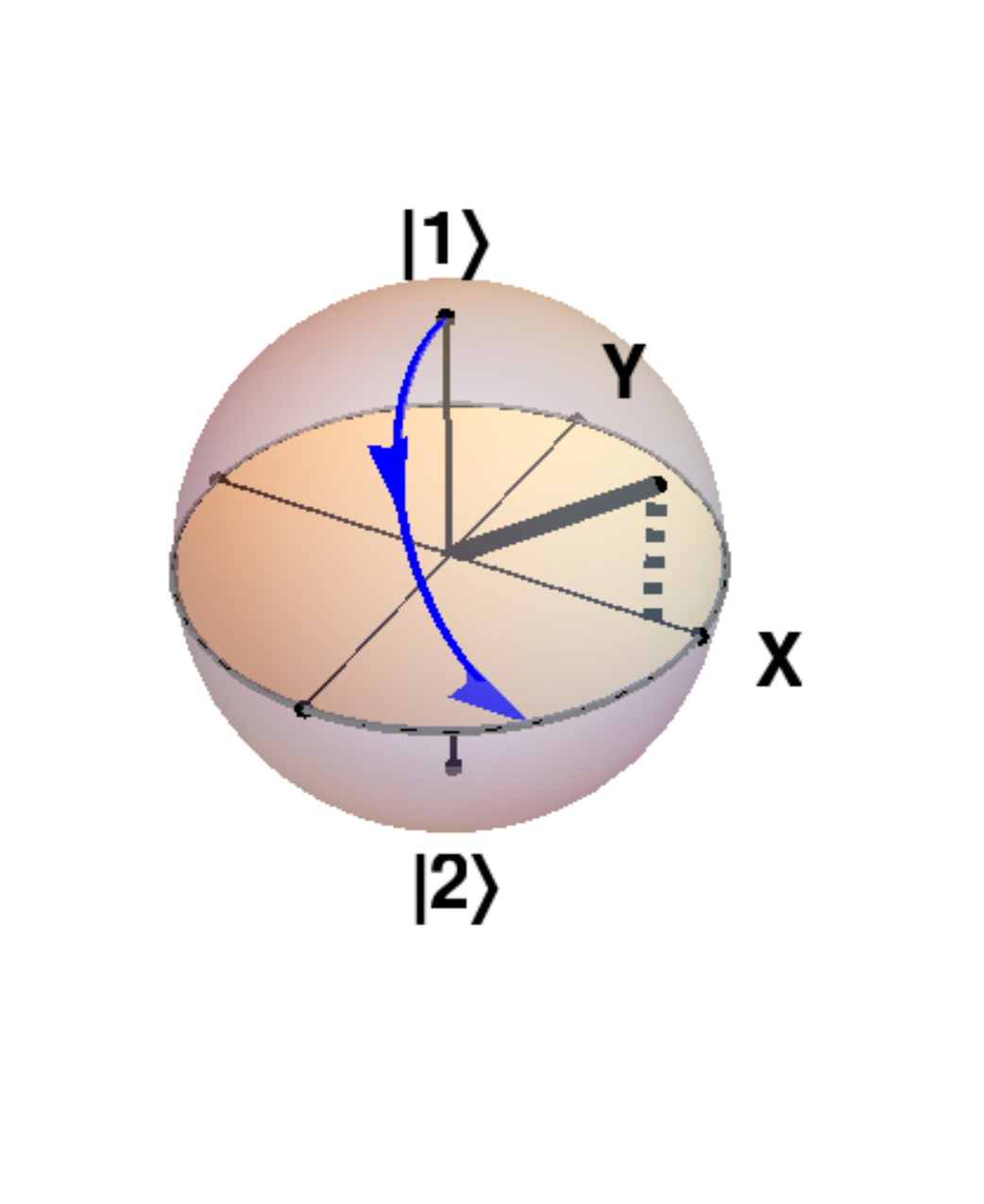}\,\includegraphics[width=0.49\linewidth]{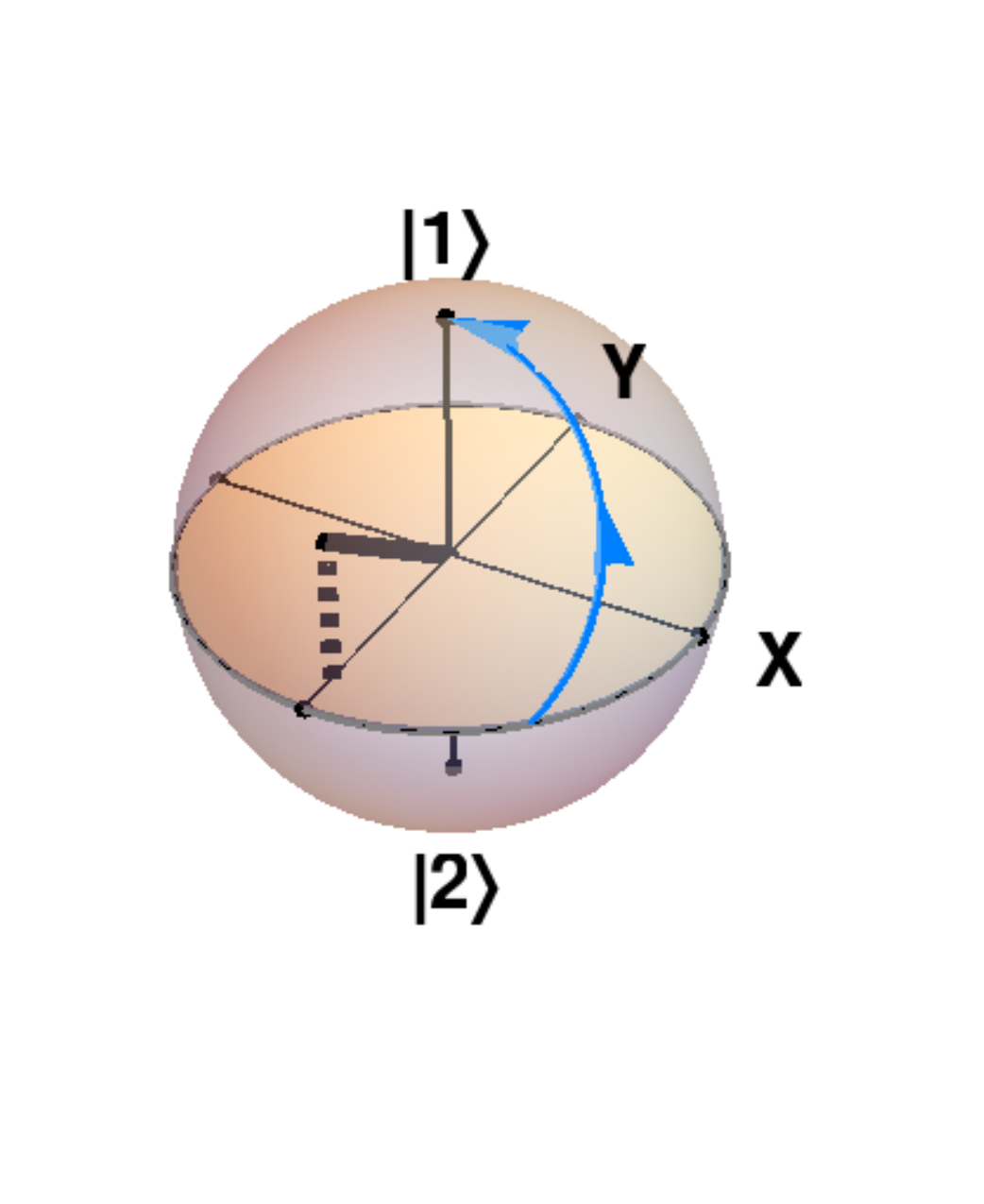}\\[0.1cm]
\end{center}
(b) Order of rotations: first $R_2({\sf T}/2)$ (left figure) and then $R_1({\sf T}/2)$ (right figure). 
\vspace*{-0.32cm}
\begin{center}
\includegraphics[width=0.49\linewidth]{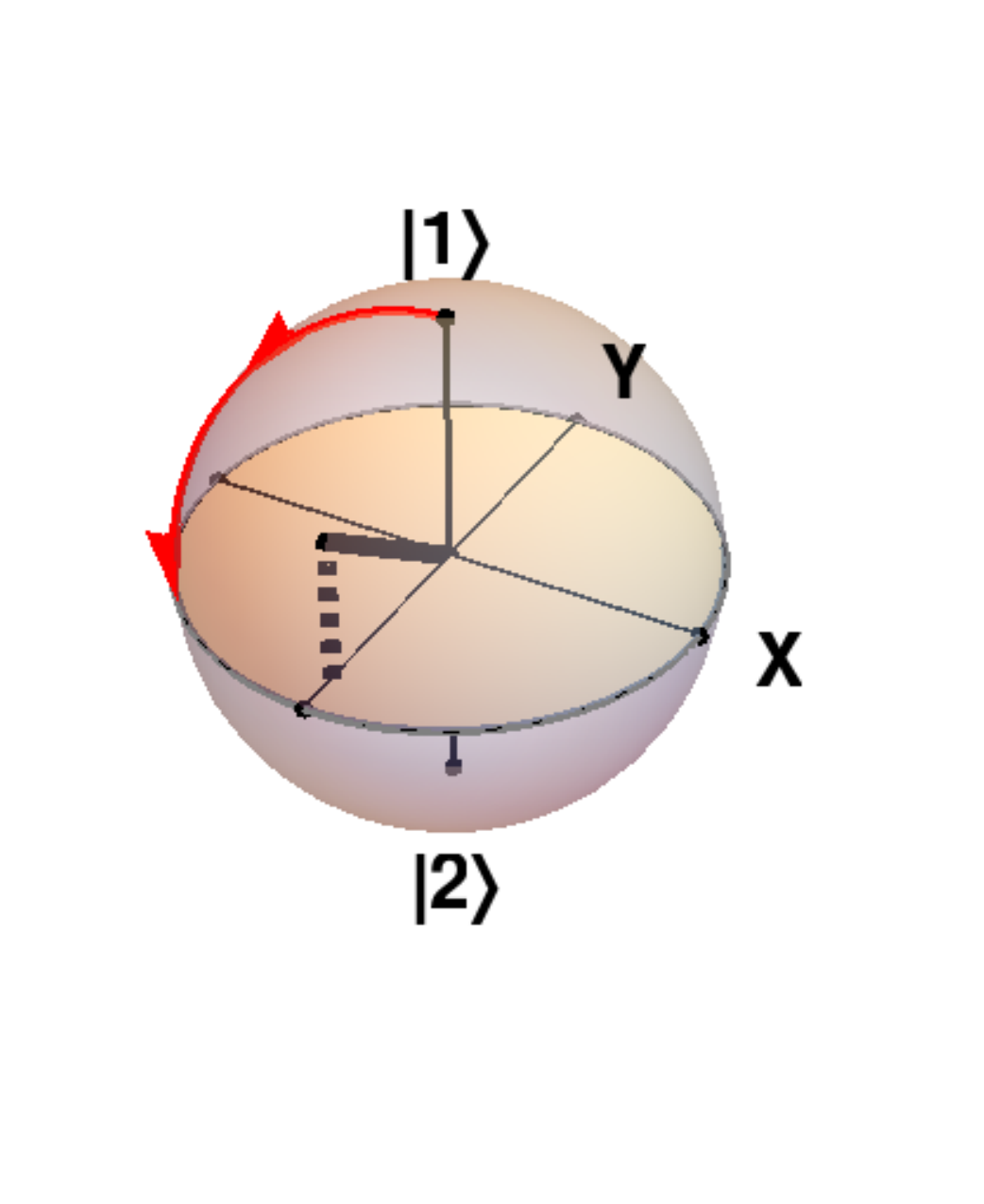}\,\includegraphics[width=0.49\linewidth]{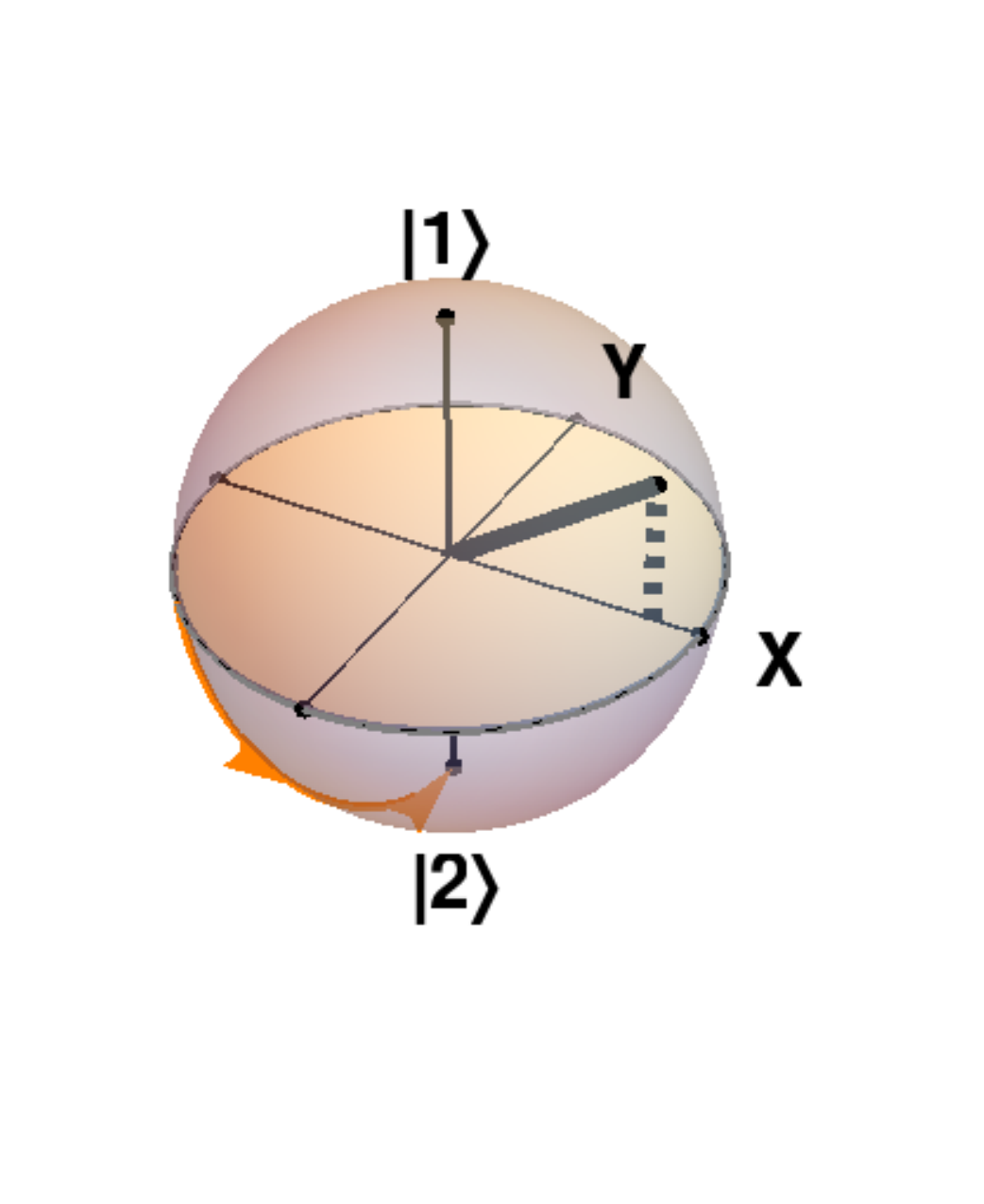}
\end{center}
\vspace*{-.8cm}
\end{minipage}
}
\caption{Simplified time-dependent model of the asymmetric ${\cal T/A}$ device with symmetry VIII: Bloch sphere explaining non time-reversal invariance, see text for details. The state trajectories are depicted in two-steps on the sphere. The rotation axes are 
also depicted. (a) The process simulates incidence from the left. The state starts and ends in $|1\ra$. (b) The process simulates incidence from the right. The state starts at $|1\ra$ and ends at $|2\ra$.  
\label{fig_t_a_simple2}}
\end{figure}

The time-evolution of this process, $\chi_\pm (t)$,
up to a phase factor may be regarded as 
two consecutive rotations $R_j=e^{-i{\beta}{\bf n}_j\cdot {\boldsymbol{\sigma}}/2}$ ($j=1,2$), with $\beta=\frac{{\sf T}}{2}\sqrt{\Omega^2+\Delta^2}$, of the two-level state on the Bloch sphere about the axes   
\beqa
{\bf n}_1&=&\frac{1}{\sqrt{\Omega^2+\Delta^2}}(\Omega,0,\Delta), 
\\
{\bf n}_2&=&\frac{1}{\sqrt{\Omega^2+\Delta^2}}(0,-\Omega,\Delta). 
\eeqa
The initial state at time $t=-{\sf T}/2$ is again $\chi_+ (-{\sf T}/2) = \chi_- (-{\sf T}/2) =\left(\begin{smallmatrix} 1\\ 0\end{smallmatrix}\right)$.
The unitary time-evolution operator to reach the final time ${\sf T}/2$ takes the form 
$e^{i\Delta {\sf T}/2}R_2 R_1$ for  incidence from the left ($\chi_+$) and
$e^{i\Delta {\sf T}/2}R_1 R_2$ for incidence from the right ($\chi_-$).  
The time ${\sf T}$ and the parameters $\Omega, \Delta$ will be fixed to reproduce the results of the full calculation with the exact model, namely, 
so that the system starts in the ground state to end either in the ground state
($\fabsq{\chi_{+} ({\sf T}/2)} = 1$)
or in the excited state by performing the rotations in one order or the reverse order
($\fabsq{\chi_{-} ({\sf T}/2)} = 0$). This gives $\Omega/\Delta = \sqrt{2}$ and ${\sf T}= 4\pi/(3 \sqrt{3} \Delta)$. It follows that ${\bf n}_1=\frac{1}{\sqrt{3}}(\sqrt{2},0,1)$ and ${\bf n}_2=\frac{1}{\sqrt{3}}(0,-\sqrt{2},1)$. 

The different outcomes can thus be understood as the result of the non-commutativity of rotations on the Bloch sphere, see    
Fig. \ref{fig_t_a_simple2}: In Fig. \ref{fig_t_a_simple2}(a), first the rotation $R_1({\sf T}/2)$ and then the rotation $R_2({\sf T}/2)$ are performed. Starting in the ground state $\ket{1}$, the system ends up  in the excited state $\ket{2}$.
In Fig. \ref{fig_t_a_simple2}(b),  first the rotation $R_2({\sf T}/2)$ and then the rotation $R_1({\sf T}/2)$ are performed:  now the system starts and ends  in the ground state $\ket{1}$. 

\begin{figure}
\begin{center}
\includegraphics[width=0.48\linewidth]{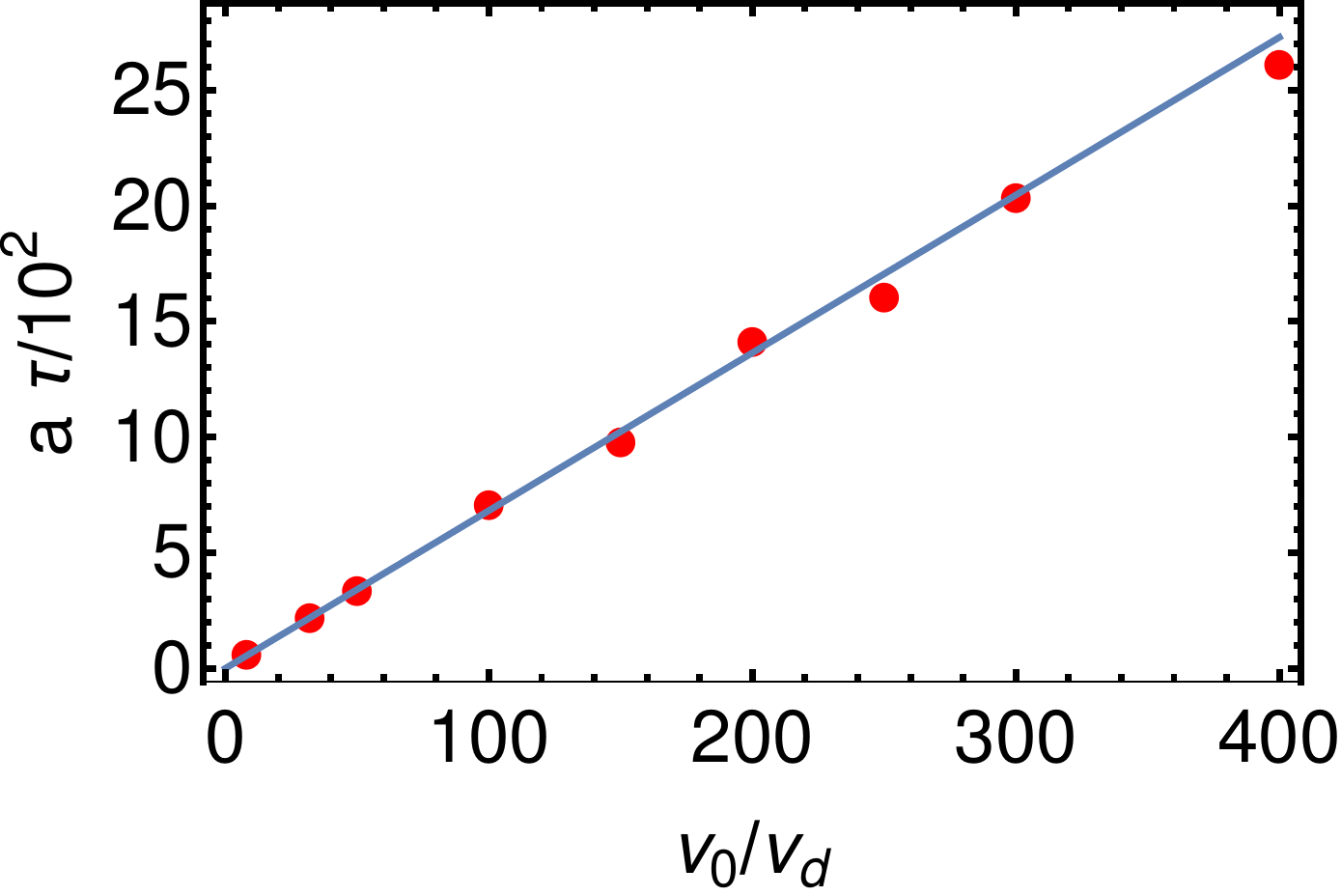}
\includegraphics[width=0.48\linewidth]{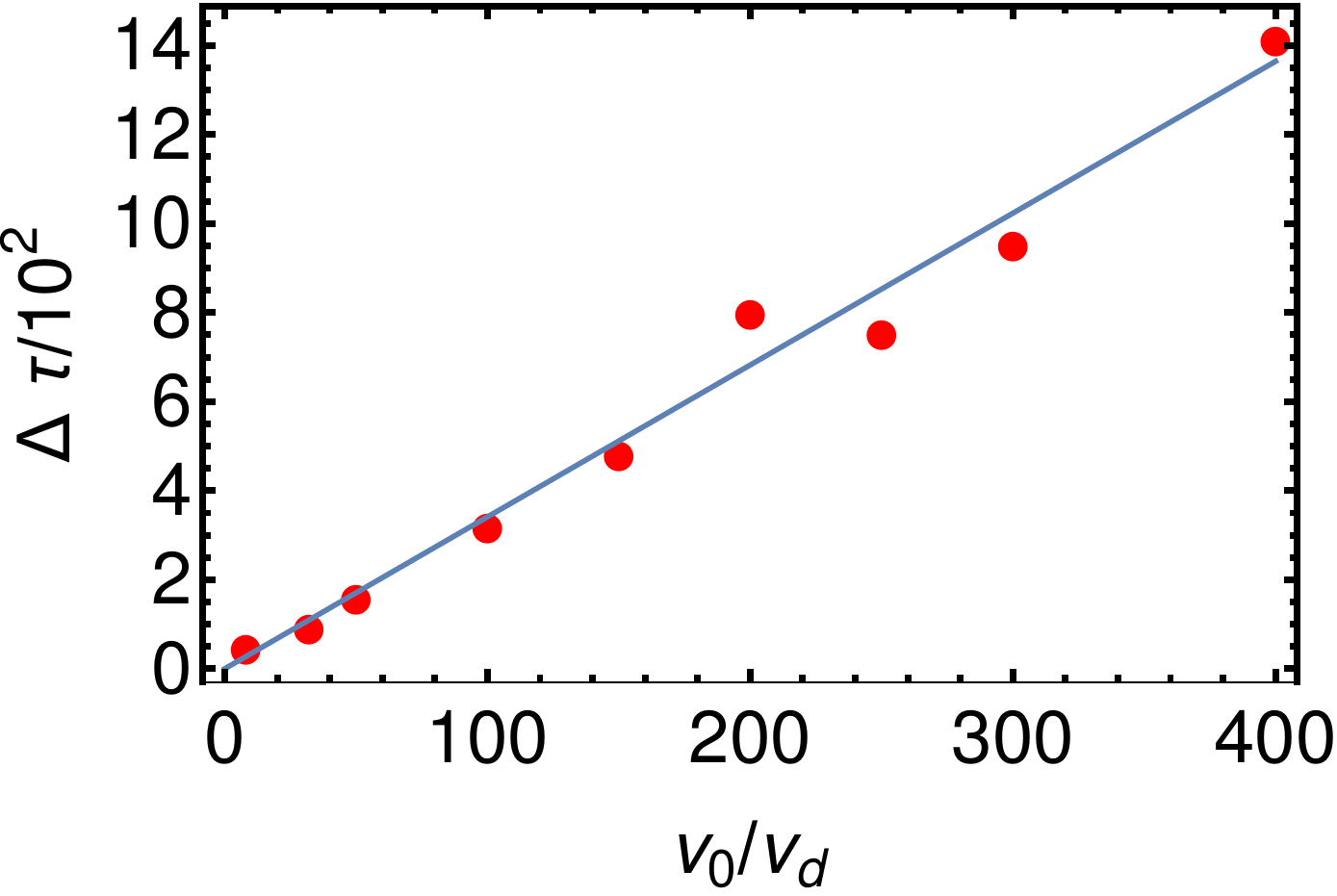}
\end{center}
\caption{Asymmetric ${\cal T/A}$ device with symmetry VIII: comparison between numerically achieved parameters (red dots) and approximated parameters (blue, solid lines) versus velocity $v_0$.
(a) Height of Rabi frequency $a$, (b) detuning $\Delta$.
\label{fig_t_a_param}}
\end{figure}

These results can be even used to approximate the parameters of the potential in the quantum setting.
As an approximation of the height $a$ we assume that the area $a \int_{-\infty}^\infty dx \, g(x) = a \sqrt{\pi} w$
is equal to $\tilde w \Omega = {{\sf T}} v_0 \Omega/2 =
v_0 \pi ({2}/{3})^{3/2}$. This results in an
approximation $a \approx \frac{v_0}{w} \sqrt{\pi}\, ({2}/{3})^{3/2}$. As an additional approximation, we
assume that $(a/\sqrt{2})/\Delta \approx {\Omega}/{\Delta} = \sqrt{2}$, so we get
$\Delta \approx a/2 \approx \frac{v_0}{2 w} \sqrt{\pi}\, ({2}/{3})^{3/2}$. A comparison between
these approximations and the numerically achieved parameters, see Fig. \ref{fig_t_a_param}, shows a good agreement
over a large velocity range.

\end{document}